\documentclass[conference]{IEEEtran}
\IEEEoverridecommandlockouts
\usepackage{cite}
\usepackage{amsmath,amssymb,amsfonts}
\usepackage{algorithmic}
\usepackage{graphicx}
\usepackage{textcomp}
\usepackage{xcolor}
\usepackage{svg}
\usepackage{epsf,exscale,times}
\usepackage[english]{babel}
\usepackage{fancyhdr}
\usepackage{siunitx}
\usepackage{subcaption}
\usepackage{algorithm}
\usepackage[acronym]{glossaries}
\usepackage{hyperref}
\usepackage{ifthen}
\usepackage{textgreek}
\usepackage{ifthen}
\usepackage{gensymb}
\usepackage{tikz}
\usepackage{textcomp} 
\usepackage{mathtools}

\usepackage{microtype}
\def\BibTeX{{\rm B\kern-.05em{\sc i\kern-.025em b}\kern-.08em
    T\kern-.1667em\lower.7ex\hbox{E}\kern-.125emX}}

\newacronym{MI}{MI}{Mutual information}
\newacronym{MINE}{MINE}{Mutual Information Neural Estimation}
\newacronym{GP}{GP}{Gaussian process}
\newacronym{MLAT}{MLAT}{multilateration}
\newacronym{TOA}{TOA}{time of arrival}
\newacronym{UWB}{UWB}{ultra wideband}
\newacronym{PMF}{PMF}{probability mass function}
\newacronym{UE}{UE}{user element}
\newacronym{CIR}{CIR}{channel impulse response}
\newacronym{PSO}{PSO}{particle swarm optimization}
\newacronym{LOS}{LOS}{line of sight}
\newacronym{SGA}{SGA}{stochastic gradient ascent}
\newacronym{BN}{BN}{Batch Normalization}
\newacronym{ELU}{ELU}{Exponential Linear Unit}
\newacronym{CPU}{CPU}{central processing unit}
\newacronym{GPU}{GPU}{graphics processing unit}
\newacronym{RAM}{RAM}{random access memory}
\newacronym{EMA}{EMA}{exponential moving average}
\newacronym{DOP}{DOP}{Dilution of precision}
\newacronym{SVD}{SVD}{singular value decomposition}
\newacronym{CRB}{CRB}{Cramér–Rao bound}
\newacronym{FIM}{FIM}{Fisher information matrix}
\newacronym{TWR}{TWR}{two-way ranging}
\newacronym{RMSE}{RMSE}{root-mean-square error}
\newacronym{AMR}{AMR}{autonomous mobile robot}
\newacronym{RNN}{RNN}{recurrent neural network}
\newacronym{LSE}{LSE}{log sum exponent}
\newacronym{RSS}{RSS}{received signal strength}
\newacronym{MVUE}{MVUE}{minimum-variance unbiased estimator}
\newacronym{PEB}{PEB}{position error bound}

\usepackage{microtype}

\begin{document}

\title{On Mutual Information Neural Estimation for Localization}
\author{\IEEEauthorblockN{Sven Hinderer, Manuel Buchfink, Bin Yang}
	\IEEEauthorblockA{\textit{Institute of Signal Processing and System Theory, University of Stuttgart, Stuttgart, Germany} \\
		firstname.lastname@iss.uni-stuttgart.de}
}

\maketitle

\noindent\textbf{Note:} This is the accepted version of the paper. 
The final version is published in the \emph{Proceedings of the IEEE 2025 15th International Conference on Indoor Positioning and Indoor Navigation (IPIN)}. 
DOI: \href{https://doi.org/10.1109/IPIN66788.2025.11213414}{10.1109/IPIN66788.2025.11213414}

\vspace{1em}

\renewcommand{\thefootnote}{}
\footnotetext{
	\textcopyright~2025 Personal use of this material is permitted.  Permission from IEEE must be obtained for all other uses, in any current or future media, including reprinting/republishing this material for advertising or promotional purposes, creating new collective works, for resale or redistribution to servers or lists, or reuse of any copyrighted component of this work in other works
}
\renewcommand{\thefootnote}{\arabic{footnote}}

\begin{abstract}
\gls{MI} is a promising candidate measure for the assessment and optimization of localization systems, as it captures nonlinear dependencies between random variables. However, the high cost of computing \gls{MI}, especially for high-dimensional problems, prohibits its application for many real-world localization systems. We evaluate an algorithm from a new class of neural \gls{MI} estimators called \gls{MINE} to approximate the \gls{MI} between the set of feasible \gls{UE} locations and the corresponding set of measurements from said \gls{UE} locations used for positioning. We apply this estimator to a simulated \gls{MLAT} system, where the true \gls{MI} for benchmarking can be approximated by Monte Carlo simulation. The estimator is experimentally evaluated w.r.t. its convergence and consistency and we investigate the usefulness of \gls{MI} for assessing simple \gls{MLAT} systems.
\end{abstract}


\section{Introduction}
The idea of using \gls{MI} \cite{shannon} to evaluate the informativeness of an experiment, in case of localization the position information contained in a measured sensor signal, goes back to \cite{mi_experiments}, where \gls{MI} was applied to experimental design.

\gls{MI} has since been utilized for various sensor placement problems that are closely related to the placement of position references in localization systems. A \gls{GP} model is proposed in \cite{MI_sensor_placement_monitoringnets}, emulating sensors with limited range and measurement uncertainty and applying \gls{MI} as a measure for the uncertainty in the \gls{GP}. The authors of \cite{placement_gp_journal} built on \cite{MI_sensor_placement_monitoringnets} and use \gls{MI} as an improved measure over the suboptimal entropy. Again, a \gls{GP} models the measurement uncertainty of range limited sensors for temperature and precipitation prediction instead of a simple disk model. The high computational cost of sensor placement optimization with \gls{MI} is handled with a greedy, iterative placement strategy and exploitation of locality in the \gls{GP} kernels.
To deal with high-dimensional problems, a lower bound of \gls{MI} is estimated in \cite{MI_subspace_bayesianopt} after mapping the samples into a lower dimensional space. Placement optimization for a chemical release problem is then performed via Bayesian optimization.
For localization, \gls{MI} was applied to evaluate \gls{RSS} based systems in \cite{MI_RSS_Monte_Carlo, MI_RSS2}, for Wi-Fi in \cite{MI_WiFi}, for fiducial marker placement in \cite{landmark_placement_navigation_cond_mi}, and to optimize the anchor placement of a \gls{TWR} \gls{UWB} indoor localization system in \cite{MI_pruning}. The \gls{MI} is computed by Monte Carlo simulation and in \cite{MI_pruning}, position reference placement optimization is performed by iterative \gls{UWB} anchor placement and massive pruning of the search space by splitting the evaluated area into smaller grids and computing the \gls{MI} only in those grids. While restricting \gls{MI} computation locally, leading to much lower computational cost, this approach loses the capability of modeling the multi modality, i.e. multiple local optima, in the likelihood function \cite{MI_pruning}. This means that while positions in the same grid with similar or identical measurements lead to decreased \gls{MI}, the same does not apply if they are in different grids.

In recent years, a new class of neural \gls{MI} estimators has been introduced \cite{MINE, MI_Rep, MI_Rep_Learning, MI_VarBounds, MI_VarLimits}. The earliest of those approaches is \gls{MINE}~\cite{MINE}, a discriminative, variational \gls{MI} estimator. \gls{MINE} shows better \gls{MI} approximations than a previous, non-parametric \textit{k}-NN \gls{MI} estimator \cite{MI_knn_kraskov,MINE}, while potentially requiring lower computational time than brute-force Monte Carlo simulation. Further, \gls{MINE} requires only sets of \gls{UE} positions and corresponding measurements for \gls{MI} estimation, and it is differentiable and therefore applicable for deep learning based \gls{MI} optimization. In this work:
\begin{itemize}
	\item We apply \gls{MINE} to the measurement model of \cite{MI_pruning}, which is equivalent to \gls{TOA} \gls{MLAT}. Therefore, the \gls{MI} between the set of evaluated \gls{UE} positions $\mathcal{X}$ and the corresponding set of measurements $\mathcal{Z}$ in a simulated \gls{MLAT} system is estimated via \gls{MINE}. Our \gls{MLAT} simulator is designed for indoor localization of \glspl{AMR}. To evaluate the \gls{MINE} estimator for this task, its \gls{MI} estimates are compared to the more precise Monte Carlo results.
	\item We investigate the correlation of \gls{MI} (estimated with Monte Carlo) and \gls{PEB} with downstream \gls{MLAT} localization performance. Even though \gls{MI} has been optimized for such systems \cite{MI_pruning, MI_pso}, an evaluation of the relation between \gls{MI} and \gls{MLAT} localization accuracy has not been done yet to the best of our knowledge.
\end{itemize}
\section{Mutual Information for Localization}
Given two discrete random variables $X$ and $Z$ with alphabets $\mathcal{X}$ and $\mathcal{Z}$ and \glspl{PMF} $p(x)=\mathrm{Pr}\{X=x\},\,x\in\mathcal{X}$ and $p(z)=\mathrm{Pr}\{Z=z\},\,z\in\mathcal{Z}$, the \gls{MI} $I(X;Z)$ between $X$ and $Z$ is defined as \cite{elements_informationtheory}
\begin{equation}
	I(X;Z)=\sum_{x\in\mathcal{X}}\sum_{z\in\mathcal{Z}}p(x,z)\mathrm{log}\left(\frac{p(x,z)}{p(x)p(z)}\right).
	\label{eq:MI_sum}
\end{equation}
Equivalently, this can be expressed as the relative entropy (KL divergence) $D_{KL}$ between the joint distribution and the product of the marginals
\begin{align}
	I(X;Z)&=D_{KL}\left( p(x,z)||p(x)p(z)\right)\notag\\
	&=\mathbb{E}_{p(x,z)}\left[\log\left(\frac{p(X,Z)}{p(X)p(Z)}\right)\right].
\end{align}
\gls{MI} measures the amount of information gained about $X$ when observing $Z$ and vice versa. Conversely, this can be seen as a decrease in uncertainty or ambiguity, which is a desirable property for localization. The above becomes clearer when formulating \gls{MI} using entropy $H(X)$ and conditional entropy $H(X|Z)$
\begin{align}
	I(X;Z)= &\,H(X)-H(X|Z)\stackrel{\text{Symmetry}}{=}H(Z)-H(Z|X)\label{eq:symm}\\
	=&-\mathbb{E}_{p(x)}\left[\log p(x)\right] + \mathbb{E}_{p(z)}\left[H(X|Z=z)\right]\nonumber\\
	=&-\sum_{x\in \mathcal{X}}p(x)\log p(x)\label{eq:entropy}\\
	&+\sum_{z\in \mathcal{Z}}p(z)\sum_{x\in\mathcal{X}}p(x|z)\log p(x|z).
	\label{eq:con_entropy}
\end{align}
For our localization problem, $\mathcal{X}$ is the set of feasible \gls{UE} positions and $\mathcal{Z}$ is the set of measurements from said positions. In principle, we are dealing with continuous random variables, as the space of \gls{UE} positions and the space of measurements are generally continuous. In this case, the sums become integrals. Since it's impossible to evaluate an infinite number of \gls{UE} positions, the feasible space of \gls{UE} positions is discretized and a finite set of measurements is drawn for said discrete locations, leading to the described discrete random variables $X$ and $Z$. We assume that the \gls{UE} height $z_0$ is fixed as normal for \glspl{AMR} and therefore $\mathcal{X}$ holds the centers of quadratic grid elements in the $xy$-plane. As evident from \eqref{eq:MI_sum}, the double sum over $\mathcal{X}$ and $\mathcal{Z}$ for \gls{MI} computation becomes very expensive for large areas discretized by small grid elements. This motivates the search for fast and reliable \gls{MI} estimators. 

\subsection{Monte Carlo \gls{MI} simulation}
In the Monte Carlo simulation \cite{MI_pruning} based on \eqref{eq:symm}, the expectation over the intractable $p(z)$ in \eqref{eq:con_entropy} is replaced by an empirical approximation by averaging over $D$ Monte Carlo samples $z_x^o$. For each \gls{UE} position $x\in\mathcal{X}$, $D$ i.i.d. noisy measurements $z_x^o$, $1\leq o\leq D$, are drawn from $p(z|x)$ to properly reflect the measurement noise distribution, hence $|\mathcal{Z}| = D|\mathcal{X}|$. The prior distribution of the \gls{UE} positions is assumed to be uniform $p(x) = 1/|\mathcal{X}|$ with entropy $H(X)=\log\left(|\mathcal{X}|\right)$ from \eqref{eq:entropy} and the Monte Carlo MI approximation $\hat{I}_{MC}$ of \eqref{eq:symm} is given with analytical $H(X)$ and approximated
\begin{equation}
	H(X|Z) = -\frac{1}{|\mathcal{X}|} \sum_{x\in\mathcal{X}}\frac{1}{D} \sum_{o=1}^D \sum_{x'\in\mathcal{X}} p(x'|z_x^o) \log p(x'|z_x^o).
	\label{eq:MC_cond_entropy}
\end{equation}
Monte Carlo simulation is expensive and requires knowledge of the likelihood $p(z|x)$ and the prior $p(x)$ for sampling and computation of the posterior $p(x|z)\stackrel{\text{Bayes'}}{=}\frac{p(z|x)p(x)}{\sum_{x\in\mathcal{X}}p(z|x)p(x)}$ and the \gls{MI}. For the simple \gls{MLAT} measurement model of \eqref{eq:measurement}, both the likelihood from \eqref{eq:likelihood} and the uniform prior are known.

\subsection{Mutual Information Neural Estimation}
The data-driven nature of \gls{MINE} allows \gls{MI} estimation also with more complex systems, e.g. when $\mathcal{Z}$ are \gls{CIR} fingerprints. In addition, \gls{MINE} is differentiable and can be used for deep learning based \gls{MI} optimization.

\gls{MINE} makes use of the Donsker-Varadhan dual representation of KL divergence \cite{MINE, KL_Donsker_Varadhan}, which gives the following lower bound on \gls{MI}
\begin{align}
	&D_{KL}(p(x,z)||p(x)p(z))\geq\notag\\
	&\quad\sup\limits_{T \in \mathcal{F}}\mathbb{E}_{p(x,z)}[T]-\mathrm{log}\left(\mathbb{E}_{p(x)p(z)}[e^T]\right),
	\label{eq:dv_bound}
\end{align}
where $\mathcal{F}$ is any class of functions $T : \Omega \rightarrow \mathbb{R}$ such that the
integrability constraints of the theorem by Donsker and Varadhan are satisfied \cite{MINE}.
\gls{MINE} then chooses a neural network called statistics network with parameters $\theta \in \Theta$ for the functions $T_\Theta:\mathcal{X} \times \mathcal{Z}\rightarrow \mathbb{R}$ and defines a new neural information measure $I_\Theta(X;Z)$
\begin{equation}
	I_\Theta(X;Z) =\sup\limits_{\theta \in \Theta} \mathbb{E}_{p(x,z)}[T_\theta]-\mathrm{log}\left(\mathbb{E}_{p(x)p(z)}[e^{T{_\theta}}]\right).
	\label{eq:dv_bound_mine}
\end{equation}

\begin{figure}[h!]
	\centering
	\includesvg[inkscapelatex=false, width=.45\textwidth]{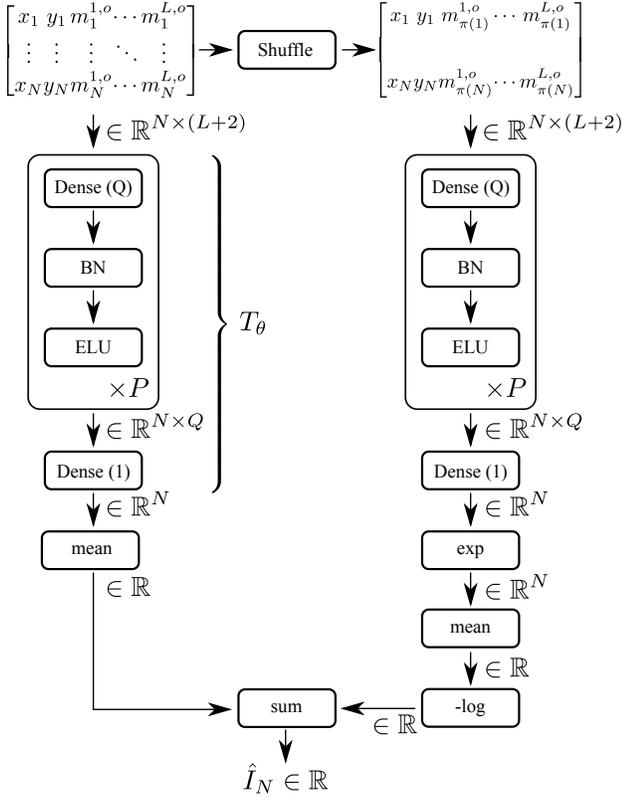}
	\caption{A mini-batch consists of $N$ stacked vectors, containing the \gls{UE} positions $(x_i, y_i)$ and corresponding measurements $(m_i^{1,o},\,\dots,\,m_i^{L,o})$. In each mini-batch update, samples from the joint distribution (left) and the marginal distribution (right) are passed through the same statistics network $T_{\theta}$. The samples from the marginals are created by shuffling the measurements along the mini-batch axis with a shuffle operator $\pi$, thus losing the correct correspondences between \gls{UE} positions and measurements. The estimated \gls{MI} $\hat{I}_N$ is then given by \eqref{eq:dv_bound_mine} and the network parameters $\theta$ are updated with \gls{SGA} until the training converges and the optimal parameters are found.}
	\label{fig:network}
\end{figure}

\gls{MI} estimation is achieved by passing mini-batches of $N$ samples from the joint distribution $p(x,z)$ and the marginals $p(x)$ and $p(z)$ through the statistics network $T_\theta$, evaluating \eqref{eq:dv_bound_mine}, updating the network parameters $\theta$ with some version of \gls{SGA}, and repeating the process until convergence, as depicted in Fig.~\ref{fig:network}. Thereby, $(x_i, y_i)\in\mathcal{X}$ is the \gls{UE} position in grid $i$ and $m_i^{j,o}$ is a corresponding measurement realization (defined in \eqref{eq:measurement}) of a position reference $j$ potentially visible from $(x_i, y_i)$. Given the set of all references $ \mathcal{L} = \{ j = 1, \ldots, L \}$, a measurement realization for grid $i$ becomes ($m_i^{1,o},\,\dots, m_i^{L,o})\in\mathcal{Z}$.
In addition, the authors of \gls{MINE} propose smoothing the denominator $\mathbb{E}\left[e^{T{_\theta}}\right]$ when computing the  derivative of the second term in \eqref{eq:dv_bound_mine} (gradient of the log) by running an \gls{EMA} for $\mathbb{E}\left[e^{T{_\theta}}\right]$ to reduce the bias in the gradients of this stochastic, mini-batch based neural \gls{MI} estimator.

It shall be noted that while \eqref{eq:dv_bound} defines a lower bound on \gls{MI}, \gls{MINE} does not give such guarantee due to the stochastic neural network training with limited number of samples $N$, possible numerical instability from the $\mathbb{E}[e^{T{_\theta}}]$ term, and the risk of overfitting. For MI estimation, we also empirically average over multiple mini-batch estimates $\mathrm{log}\left(\mathbb{\hat{E}}\left[e^{T_{\theta}}\right]\right)$ with $N$ samples per mini-batch. This leads to an overestimation of \eqref{eq:dv_bound} because of the concavity of the logarithm \cite{mi_var_bounds, condMINE}, since Jensen's inequality states $\mathrm{log}\left(\mathbb{E}\left[e^{T_{\theta}}\right]\right){\geq}\,\mathbb{E}_{batch}\left[\,\mathrm{log}\left(\mathbb{\hat{E}}\left[e^{T_{\theta}}\right]\right)\right]$ , where $\mathbb{E}_{batch}$ denotes the empirical average over mini-batches. Further, \gls{MINE}'s consistency assumption relies on the universal approximation theorem, which gives no practical guarantees to find an optimal $T_{\theta}$ that bounds the \gls{MI} approximation error. 
\section{Position Error Bound for \gls{MLAT}}
\gls{DOP} and the related \gls{PEB} are cheap to compute and can be used as localization metrics to evaluate \gls{MLAT} systems. We will compare \gls{PEB} with \gls{MI} in the experiments.
For a \gls{TOA} \gls{MLAT} system with  zero mean, i.i.d. Gaussian noise with standard deviation $\sigma_r$ for all references, the achievable \gls{PEB} of an efficient \gls{MVUE} for an evaluated \gls{UE} position $\underline{q}_i= \left[x_i, y_i, z_0\right]^T$ from $\mathcal{X} = \left\{ \underline{q}_i  \;\middle|\; 1\leq i\leq K \right\}$ is \cite{rmse_dop}
\begin{align}
	\mathrm{PEB}(\underline{q}_i) = \sigma_r\sqrt{\mathrm{tr}\left(\mathbf{H}_i^T\mathbf{H}_i\right)^{-1}},
	\label{eq:rmse_ideal}
\end{align}
where tr($\mathbf{H}_i$) denotes the trace of matrix $\mathbf{H}_i$, $\mathbf{H}_i^{\textrm{-1}}$ its inverse and  $\mathbf{J}_i=\frac{1}{\sigma^2_r}\mathbf{H}_i^T\mathbf{H}_i$ is the~\gls{FIM}. The matrix $\mathbf{H}_i$ contains the geometric information of the reference placement and is computed with the locations of the $\mathcal{R}_i \subseteq \mathcal{L}$,  position references $\underline{p}_j=\left[x_j, y_j, z_j\right]^T,\,j\in \mathcal{R}_i$, visible from $\underline{q}_i$ as
\begin{align}
	\mathbf{H}_i = \left[\underline{h}_{i,j}\right]^T_{j\in\mathcal{R}_i} \quad\mathrm{with}\quad
	\underline{h}_{i,j} = \frac{\underline{q}_i-\underline{p}_j}{\lVert\underline{q}_i-\underline{p}_j \rVert}.
\end{align}
Since we know the \gls{UE} height $z_0$, we compute 2D \gls{PEB} with only the first (for the standard deviation $\sigma_x$ in $x$) and the second (for the standard deviation $\sigma_y$ in $y$) diagonal elements in the trace in \eqref{eq:rmse_ideal}.
\section{Simulation environment}
Our \gls{MLAT} system simulator is based on our previous work \cite{ipin_sven}, where we introduced an advanced \gls{PSO} algorithm to optimize the position reference placement of our novel radar-based indoor localization system.
A feasible \gls{UE} position vector $\underline{q}_i= \left[x_i, y_i, z_0\right]^T$ in our localization system is defined as the center of one of $K$ grid elements in the $xy$-plane. For a \gls{UE} position to be considered feasible, it has to be inside the evaluated room. The known height $z_0$ of the \gls{UE} will be dropped later for \gls{MI} and \gls{UE} position estimations. The room is equipped with a number of $L=|\mathcal{L}|$ identifiable position references used for localization. One of $o=1,\,\dots, D$ measurement realizations $m_i^o = \left(m_i^{j,o}\right)_{j\in\mathcal{R}_i}\in\mathcal{Z}$ per \gls{UE} position $\underline{q}_i$ contains $R_i=|\mathcal{R}_i|$ noisy, absolute range measurements $m_{i}^{j,o}$, with
\begin{equation}
	m_{i}^{j,o} = r_i^j + n_{i}^{j,o},
	\label{eq:measurement}
\end{equation}
where $r_i^j=\|\underline{q}_i-\underline{p}_j\|$ is the true 3D distance between the \gls{UE} at $\underline{q}_i$ and the position reference at $\underline{p}_j$. We assume that the noise samples $n_{i}^{j,o}$ for each reference $j$ are i.i.d. and Gaussian or uniformly distributed noise. The likelihood $p(m_{i'}^{o}| \underline{q}_i)$ for a measurement realization $m_{i'}^{o}$ from \gls{UE} grid element $i'$  given \gls{UE} position $\underline{q}_i$ is defined with
\begin{align}
	p_{\text{Gauss}}(m_{i'}^{j,o} \mid \underline{q}_i) &= \frac{1}{\sqrt{2\pi \sigma_r^2}} 
	\exp \left( -\frac{\| m_{i'}^{j,o} - r_i^j \|^2}{2\sigma_r^2} \right) \label{eq:gauss} \\[1ex]
	p_{\text{Uni}}(m_{i'}^{j,o} \mid \underline{q}_i) &=
	\begin{cases}
		\frac{1}{2\sqrt{3} \sigma_r}, & \text{if } \left| m_{i'}^{j,o} - r_i^j \right| \leq \sqrt{3} \sigma_r \\
		0, & \text{otherwise}
	\end{cases} \label{eq:uni} \\[1ex]
	p(m_{i'}^o \mid \underline{q}_i) &=
	\begin{cases}
		\prod_{j\in\mathcal{R}_i} p(m_{i'}^{j,o} \mid \underline{q}_i),& \text{if defined} \\
		0,& \text{otherwise},
	\end{cases} \label{eq:likelihood}
\end{align}
where \eqref{eq:likelihood} equals zero if not exactly the same $\mathcal{R}_i$ references that should be visible from $\underline{q}_i$ are contained in $m_{i'}^o$. We apply a disk model to simulate a limited 2D sensing range of the position references in the $xy$-plane and consider them detectable, if the \gls{UE} is in the sensing range of the position reference and if there is \gls{LOS} between the \gls{UE} and the reference. To replicate this behavior with the fixed-length input of \gls{MINE} containing all $L$ references, we set the measurements $m_{i}^{j,o}$ of the $\mathcal{L}\setminus\mathcal{R}_{i}$ undetected references to $-1$. Such masking is safely applicable if the mask value does not naturally occur in the data and generally does not influence the neural network results, as the network learns to treat it as missing value \cite{chollet_dl_python}.
\section{Experiments}
\subsection{Simulated environments}
We evaluate the \gls{MINE} estimator for a small quadratic, i.e. convex room of size (4, 4, 3)~\si{m} with 4 position references and $K=400$ grid elements, where \gls{LOS} to all references exists from every \gls{UE} position. Further, a larger L-shaped room of size (10, 10, 3)~\si{m} with 8 references and $K=1875$ grid elements is considered, where \gls{LOS} is not always given, as the room is non-convex, and visibility is also limited by the sensing range of references. The rooms are discretized by grid elements of size (0.2, 0.2)~\si{m}. The \gls{UE} height is $z_0=$~\SI{10}{cm}. To ensure stable training, we found it important to set the grid size not much larger than the noise standard deviation, likely due to overfitting. The position references have a 2D sensing range of \SI{7.4}{m}. All position references are installed at a height of $z_j=z_1=2.5$~\si{m}. They have a minimum spacing of \SI{0.5}{m} to each other and a minimum distance of \SI{0.5}{m} to the nearest wall. For modeling the distributions of the range measurement noise samples $n_i^{j,o}$ with standard deviation $\sigma_r = 0.2$~\si{m}, we draw $n_{i}^{j,o}$ either from the Gaussian distribution $n_{i}^{j,o} \sim \mathcal{N}(0, \sigma_r^2)$  or from the uniform distribution $n_i^{j,o} \sim \mathcal{U}\left(-\sqrt{3}\sigma_r,\ \sqrt{3}\sigma_r\right)$. All Monte Carlo simulations for \gls{MI} benchmarking and all \gls{MLAT} simulations are run with $D=1000$ i.i.d. measurements per grid element.
 
\subsection{Statistics network architecture and training}
For the statistics network $T_{\theta}$, we choose a stack of $P$ dense layers with $Q$ neurons, \gls{BN} \cite{BatchNorm} and \gls{ELU} activation \cite{ELU_act}. The network architecture is depicted in Fig.~\ref{fig:network}.

Three different model sizes are investigated with $Q=32$, $P=2$ for the small model, $Q=128$, $P=3$ for the medium model and $Q=256$, $P=4$ for the large model. To train the networks, we use the Adam optimizer \cite{adam} with default settings and an exponentially decaying learning rate $lr$ defined by the schedule
\begin{equation}
	lr = 10^{-3} \cdot 0.98^{ \frac{\mathrm{epoch}}{2000}}.
\end{equation}
We train on \gls{CPU} instead of \gls{GPU}, which has shown to make little difference w.r.t. training times given our models and datasets. This allows us to train many \gls{MINE} models in parallel for evaluation and also gives us access to \gls{RAM}, thus enabling training with large mini-batches of size $N=K=|\mathcal{X}|$. Using large mini-batches has shown to improve stability and convergence for training. In our experiments, we define an epoch as training on a new set of independent measurements per grid element, i.e. a mini-batch is equivalent to an epoch, and we train with 500.000 epochs for the small (quadratic) room and with 700.000 epochs for the large (L-shaped) room. We don't use the smoothed \gls{EMA} of $\mathbb{E}\left[e^{T{_\theta}}\right]$. Although it initially improved convergence, it also resulted in numerical instability during training. We suspect this is caused by explicitly having to compute $\mathbb{E}\left[e^{T{_\theta}}\right]$ with the \gls{EMA} gradient updates, which can otherwise be circumvented by using the log-sum-exp trick in \eqref{eq:dv_bound_mine}.

To estimate the \gls{MI} $\hat{I}_{MINE}$ with \gls{MINE}, we average the 20.000 \gls{MINE} estimates $\hat{I}_N$ from the most recent epochs. Larger averaging intervals lead to better \gls{MI} estimates as more samples are used in the estimation, but they also slow down convergence, since older \gls{MI} estimates from a less trained \gls{MINE} model are included in the \gls{MI} estimation.
\subsection{Relationship between \gls{MI} and \gls{MLAT} performance.}
It was argued in \cite{MI_pso} that the \gls{MI} between \gls{UE} positions and measurements does not properly reflect \gls{MLAT} performance, as \gls{MI} still gives us some information, even when a \gls{UE} receives signals from only one position reference. In such scenarios, \gls{UE} localization will fail. The authors therefore advocate for using a metric based on the determinant of the \gls{FIM} inverse, similar to \gls{DOP} or \gls{PEB} \cite{rmse_dop}. We simply solve this issue with \gls{MI} by setting a coverage boundary condition for the position reference placement, such that a minimum of 4 references must be visible from all feasible \gls{UE} positions \cite{ipin_sven} and we ensure that the \gls{FIM} in \eqref{eq:rmse_ideal} is not singular. In addition, we evaluate localization performance with a nonlinear least squares \gls{MLAT} algorithm for single snapshots w.r.t. \gls{MI} and \gls{PEB} to see if \gls{MI} is a viable measure for \gls{MLAT} performance.
\section{Results}
\subsection{Convergence of MINE}
\begin{figure}[h!]
	\centering
	\begin{subfigure}[b]{.40\textwidth}\
		\centering
		\includegraphics[width=\textwidth]
		{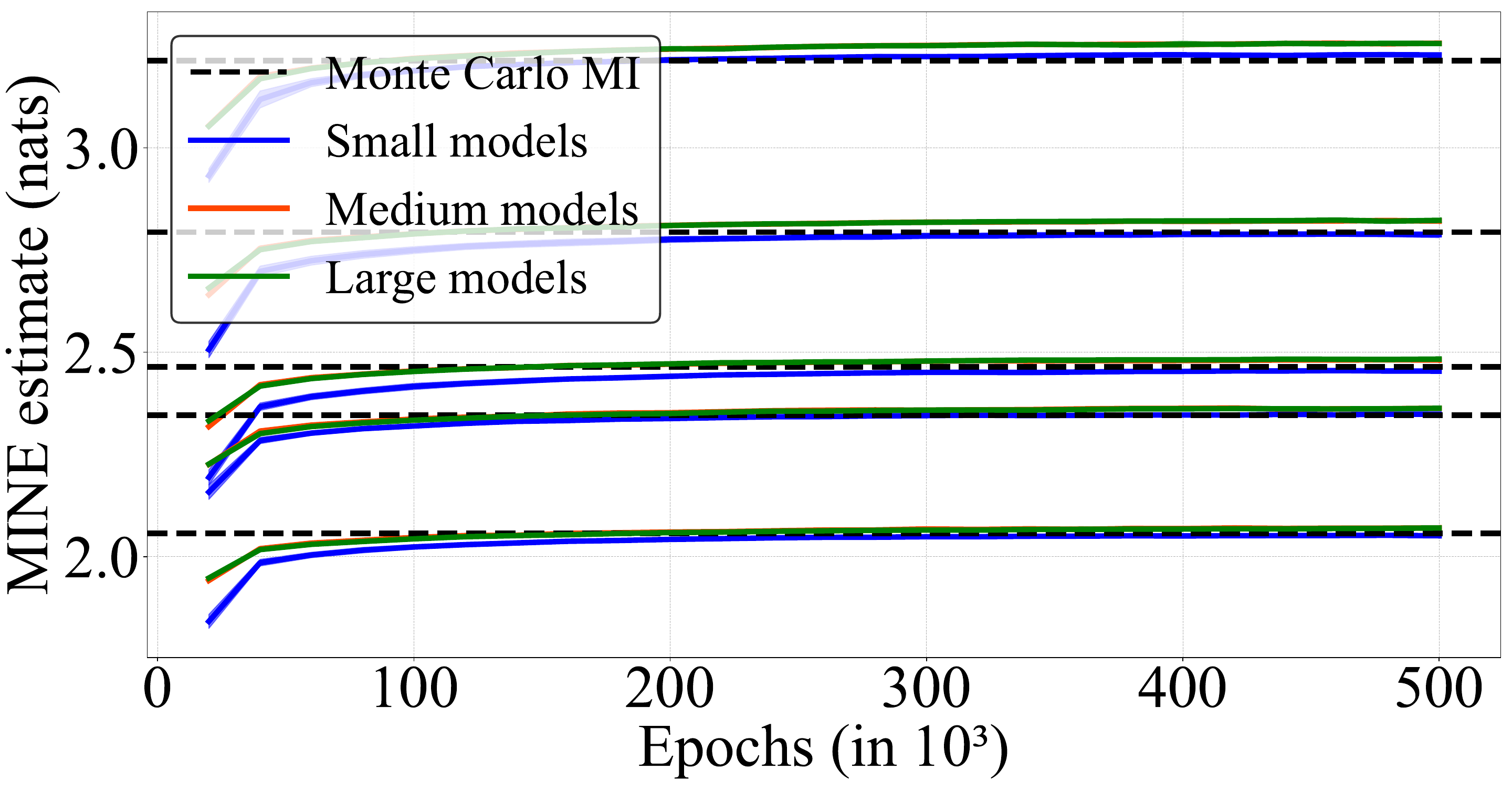}
		\caption{4x4 m room, Gaussian noise.}
		\label{fig:convergence_4x4_gauss}
	\end{subfigure}
	\vfill
	\begin{subfigure}[b]{.40\textwidth}
		\centering
		\includegraphics[width=\textwidth]{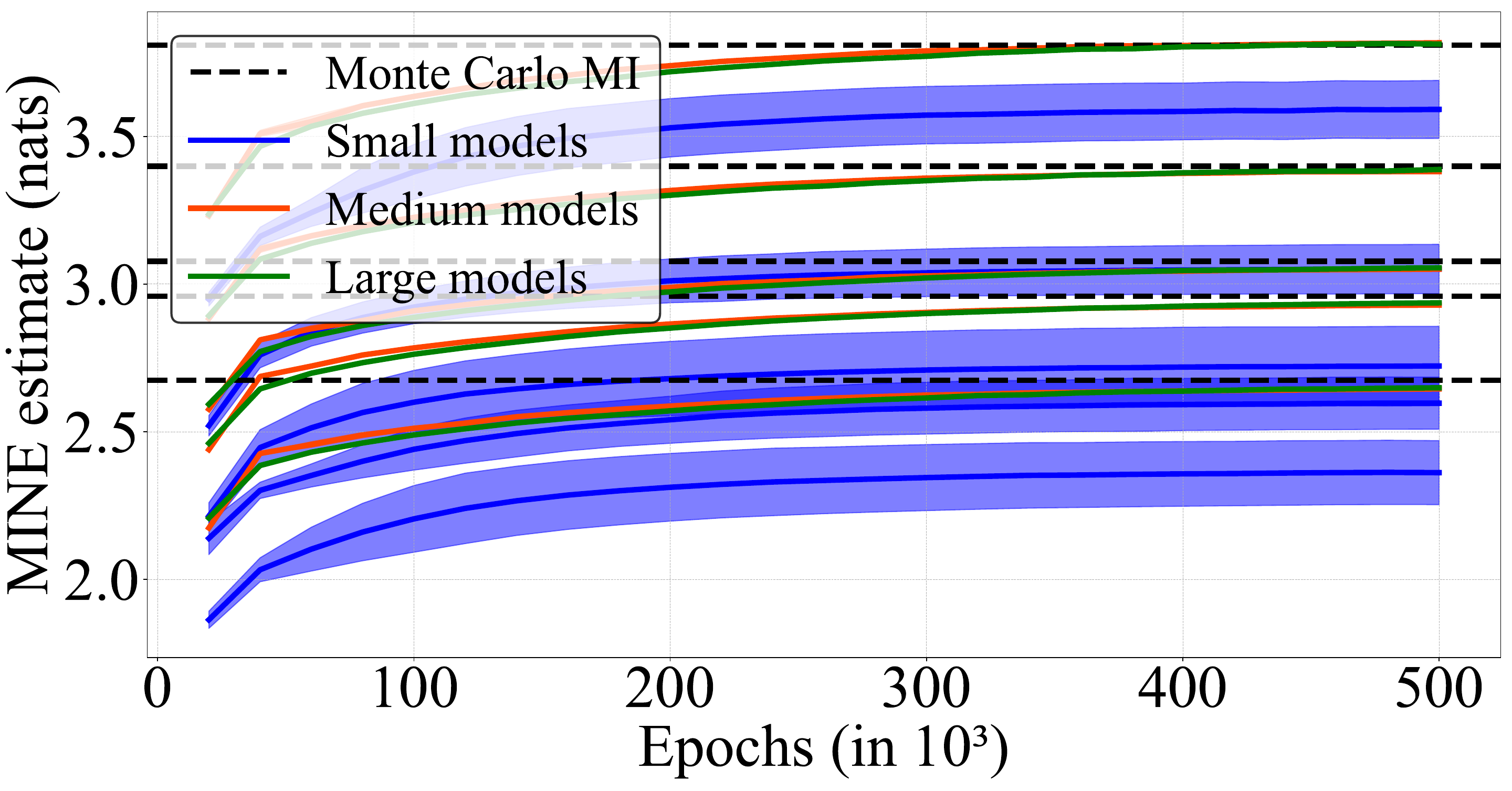}
		\caption{4x4 m room, uniform noise.}
		\label{fig:convergence_4x4_uni}
	\end{subfigure}
	\vfill	\begin{subfigure}[b]{.40\textwidth}\
		\centering
		\includegraphics[width=\textwidth]
		{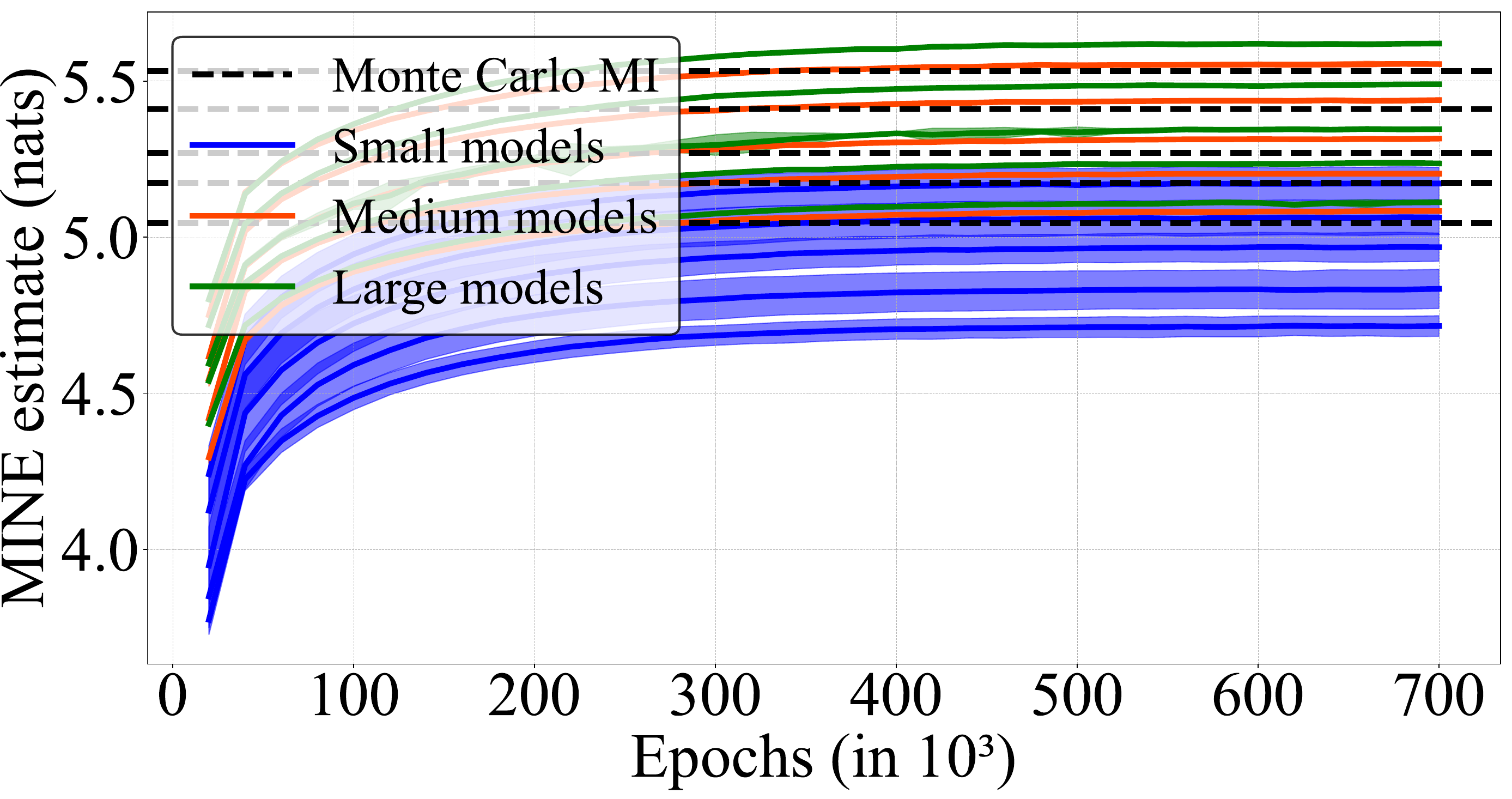}
		\caption{10x10 m room, Gaussian noise.}
		\label{fig:convergence_10x10_Gauss}
	\end{subfigure}
	\vfill
	\begin{subfigure}[b]{.40\textwidth}
		\centering
		\includegraphics[width=\textwidth]{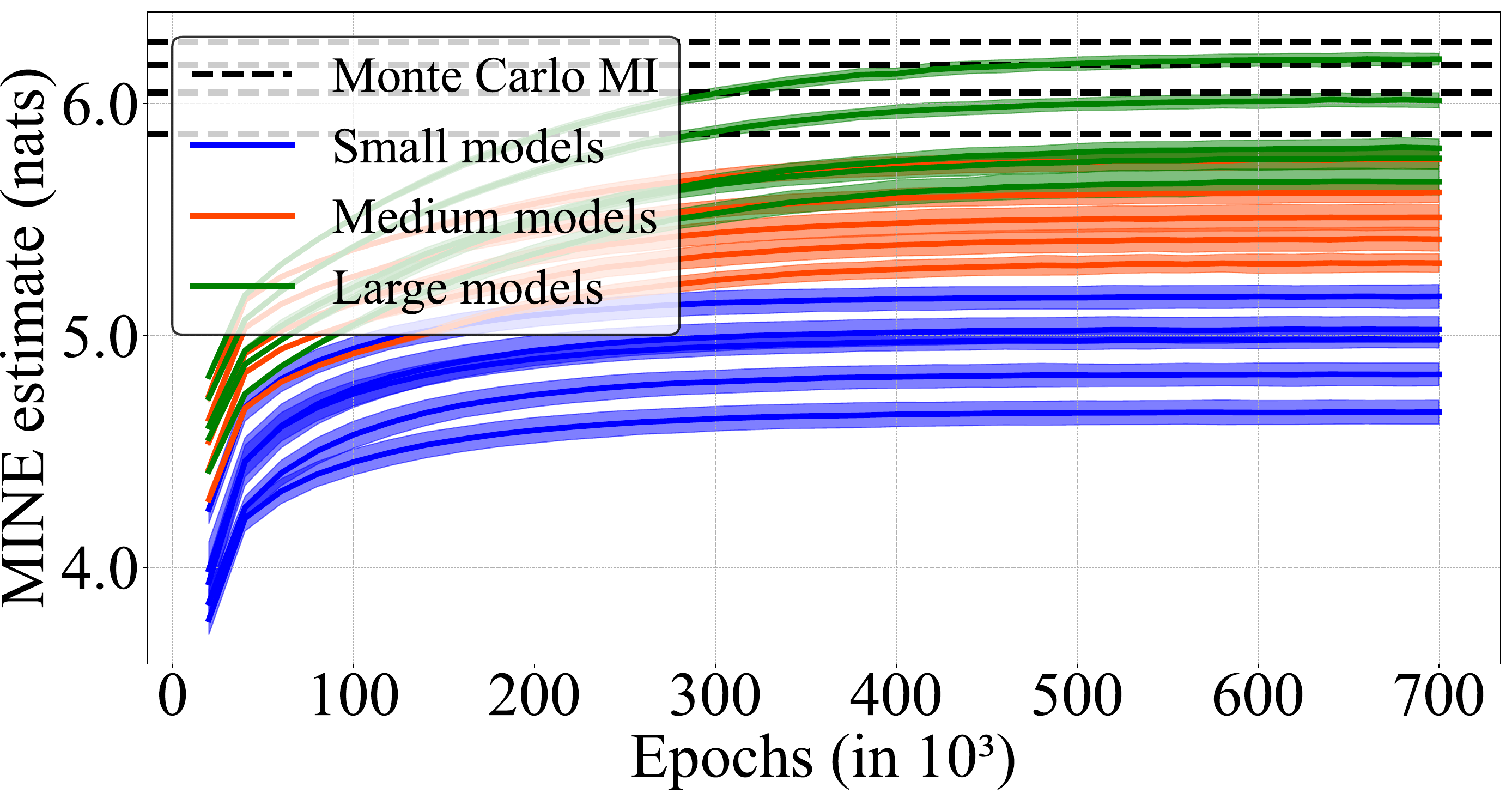}
		\caption{10x10 m room, uniform noise.}
		\label{fig:convergence_10x10_uni}
	\end{subfigure}
	\caption{Convergence of \gls{MINE} models of different size for five position reference placements per room. Shown is the desired \gls{MI} approximated by Monte Carlo simulation. Each curve gives the mean and standard deviation of 10 independent, equally sized \gls{MINE} models trained on one of the five position reference placements per room.}
	\label{fig:convergence}
\end{figure}

For the evaluation of the convergence of \gls{MINE}, we train 10 independent \gls{MINE} models per reference placement. This is repeated for each model size and for five different reference placements per room. The learning curves of the \gls{MINE} models are shown in Fig.~\ref{fig:convergence}, where each curve gives the mean and standard deviation of the 10 respective, equally sized \gls{MINE} models per reference placement and room.

For this simulated \gls{MLAT} system, \gls{MINE} converges smoothly most times, but not always to the Monte Carlo \gls{MI}. The best convergence is given with the Gaussian noise in the small room in Fig.~\ref{fig:convergence_4x4_gauss}, but \gls{MI} is consistently overestimated. Since the smallest models approximate the Monte Carlo \gls{MI} the closest in this simple environment, we suspect mild overfitting and overestimation due to limited grid resolution and mini-batch size. Using the same room and uniform noise depicted in Fig.~\ref{fig:convergence_4x4_uni}, the small models underestimate the \gls{MI}. They also estimate different \gls{MI} values for the same reference placements, seen in the large standard deviations of the blue curves. Medium and large models give accurate \gls{MI} estimates, but require long training.
For the large room with Gaussian noise in Fig.~\ref{fig:convergence_10x10_Gauss}, the \gls{MINE} behavior is similar. Small models fail again with \gls{MI} estimation. Medium and large models slightly overestimate the \gls{MI} like in the small room with Gaussian noise in Fig.~\ref{fig:convergence_4x4_gauss}. The large room with uniformly distributed noise in Fig.~\ref{fig:convergence_10x10_uni} shows the worst \gls{MI} estimates. All models underestimate the \gls{MI} and have large standard deviations. Discriminative, neural \gls{MI} estimators like \gls{MINE} are known to have trouble with estimating larger \gls{MI} values \cite{MI_VarLimits}, which might influence this result. Large models give the closest \gls{MI} estimates and larger statistics networks in combination with smaller grid sizes and longer training could potentially fill the remaining gap to reach the Monte Carlo \gls{MI}. Overall, \gls{MINE} shows good convergence, but stable training and correct \gls{MI} estimates require careful tuning of simulation-, model- and hyperparameters, and also long model training.

\subsection{Consistency of \gls{MINE}}
In this experiment, we select a set of 50 (the previous 5 and 45 new) reference placements per room that span a larger range of \gls{MI} values. For each reference placement, one \gls{MINE} model is trained. The Pearson correlation coefficient $\rho = \frac{\mathrm{cov}(\hat{I}_{MINE}, \hat{I}_{MC})}{\sigma_{\hat{I}_{MINE}} \sigma_{\hat{I}_{MC}}}$ between the Monte Carlo \glspl{MI} $\hat{I}_{MC}$ and the \gls{MINE} estimates $\hat{I}_{MINE}$ is computed. In case of perfect (linear) correlation of the \gls{MINE} estimates with the Monte Carlo estimates, $\rho$ would be $\pm1$.
\begin{figure}[t!]
	\centering
	\begin{subfigure}[b]{.24\textwidth}\
		\centering
		\includegraphics[width=\textwidth]
		{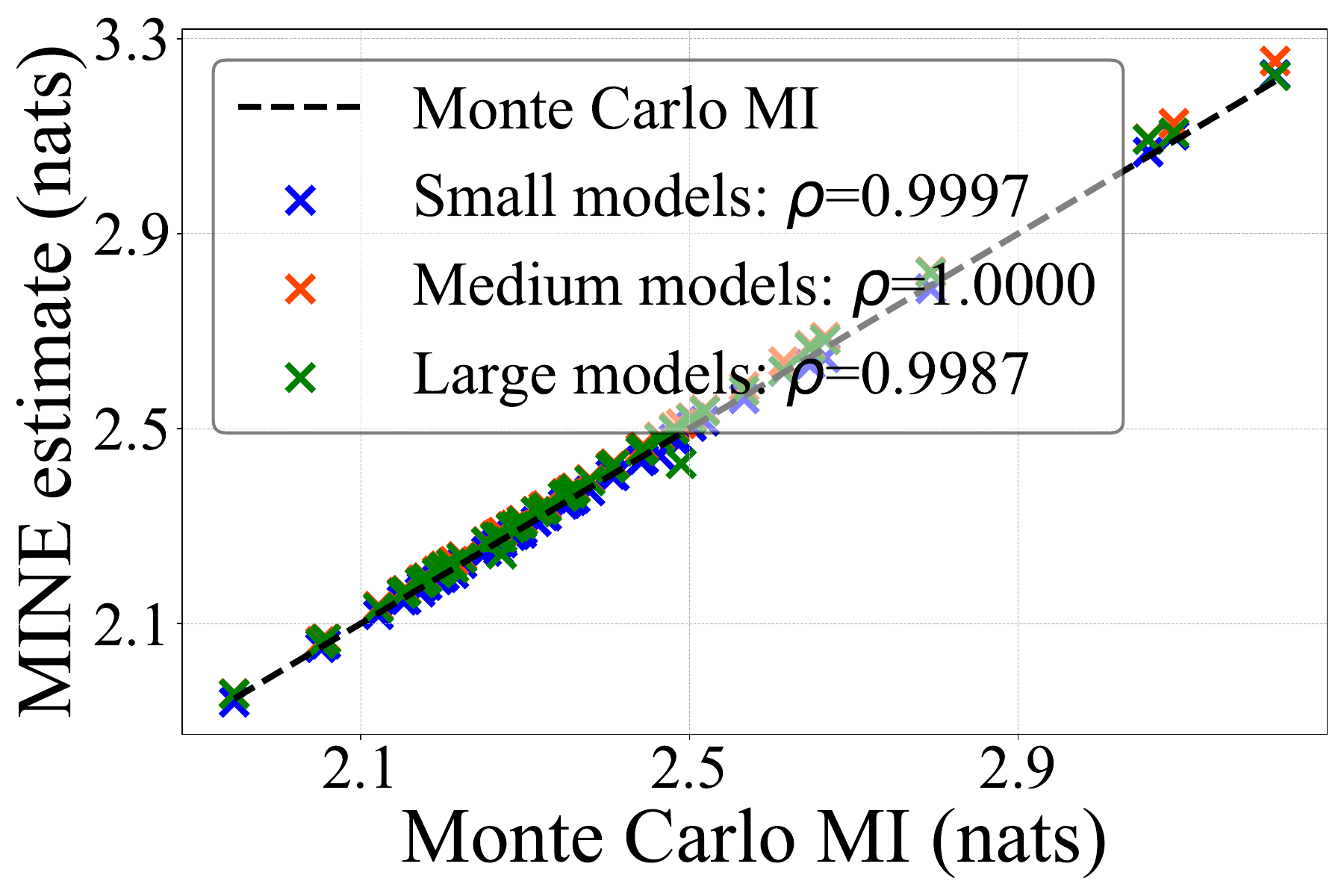}
		\caption{4x4 m room, Gaussian noise.}
		\label{fig:corr_4x4_gauss}
	\end{subfigure}
	\hfill
	\begin{subfigure}[b]{.24\textwidth}
		\centering
		\includegraphics[width=\textwidth]{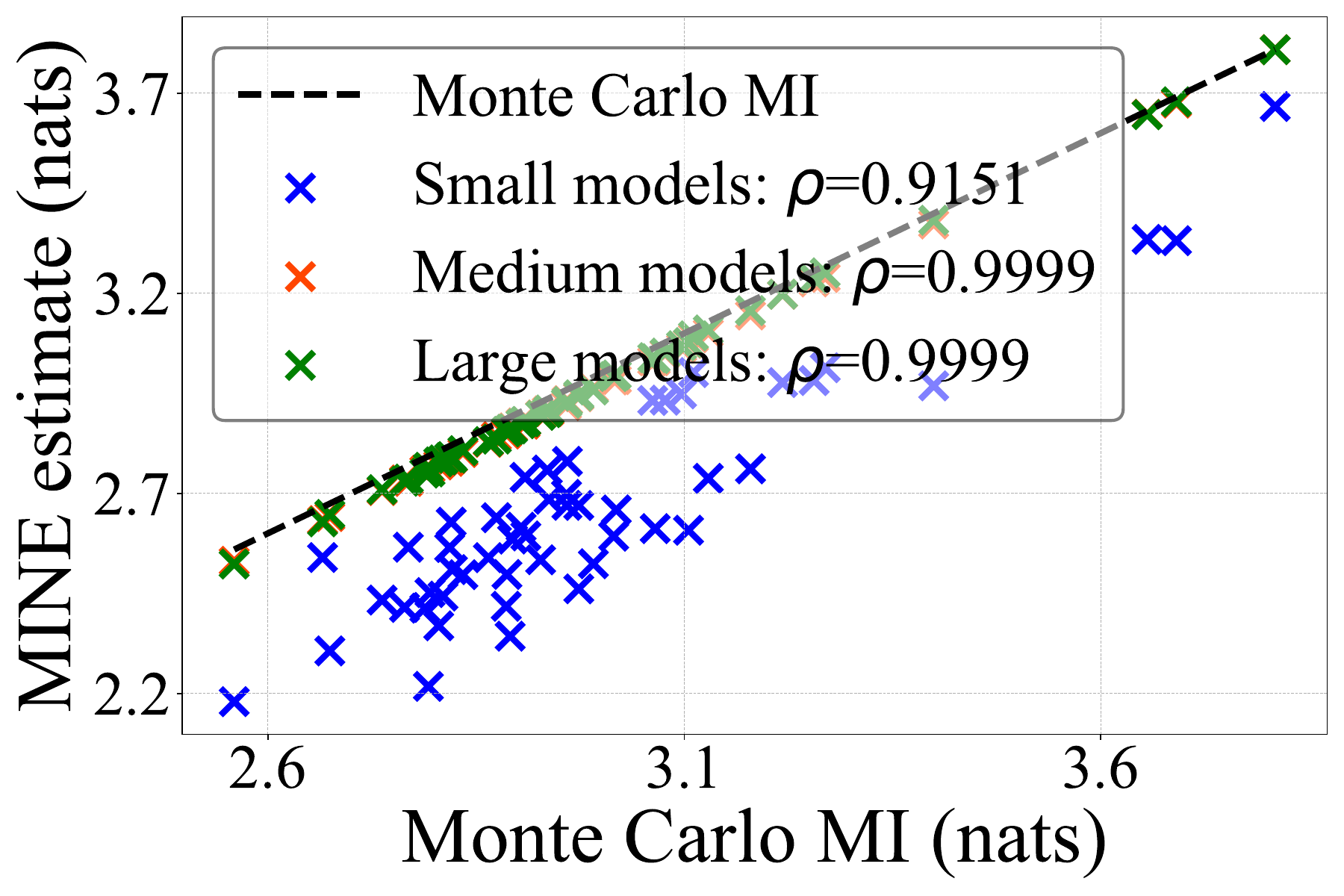}
		\caption{4x4 m room, uniform noise.}
		\label{fig:corr_4x4_uni}
	\end{subfigure}
	\vfill
		\begin{subfigure}[b]{.24\textwidth}\
		\centering
		\includegraphics[width=\textwidth]
		{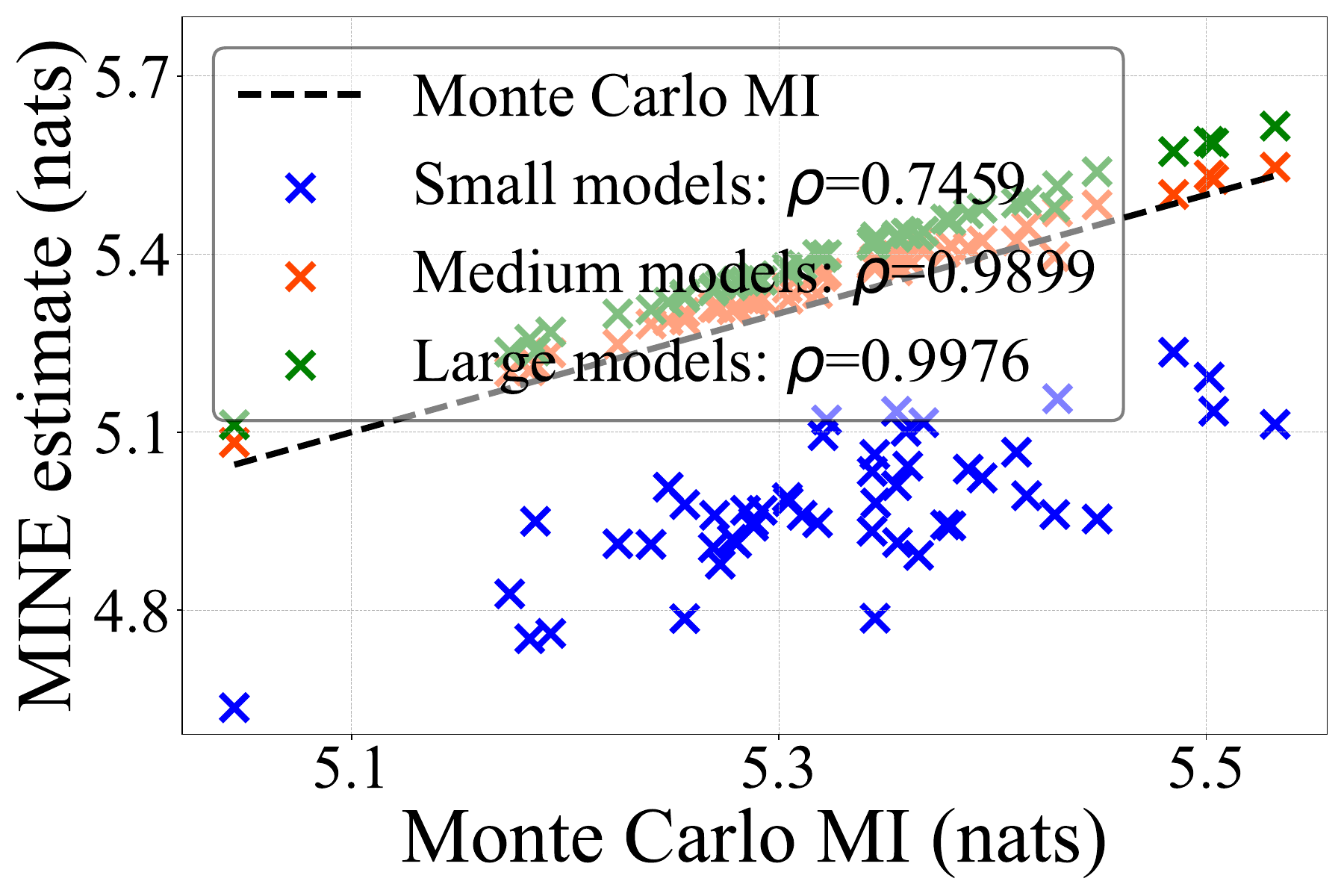}
		\caption{10x10 m room, Gaussian noise.}
		\label{fig:corr_10x10_gauss}
	\end{subfigure}
	\hfill
	\begin{subfigure}[b]{.24\textwidth}
		\centering
		\includegraphics[width=\textwidth]{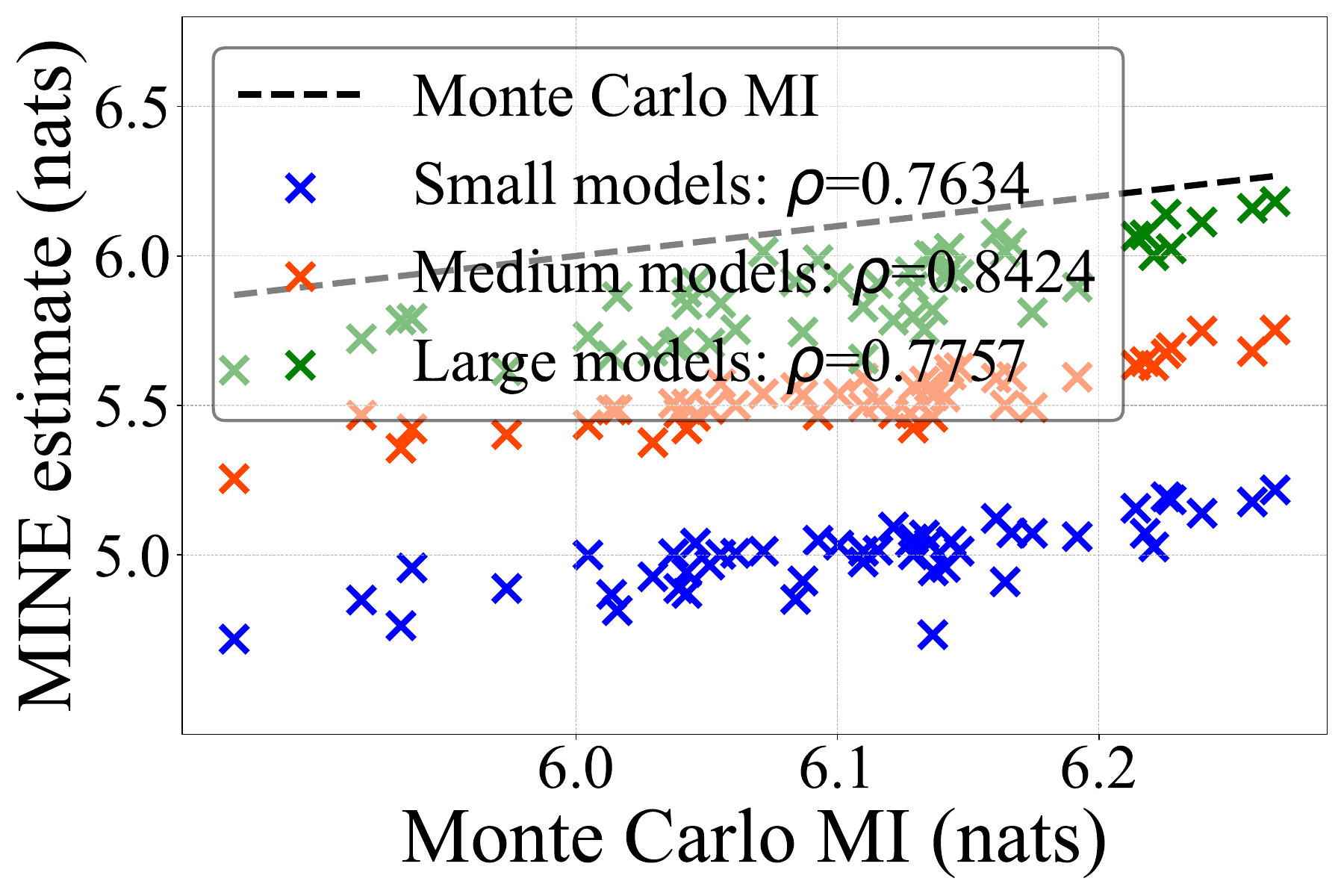}
		\caption{10x10 m room, uniform noise.}
		\label{fig:corr_10x10_uni}
	\end{subfigure}
	\caption{\gls{MINE} consistency check with 50 different position reference placements per room. Close approximation of the Monte Carlo \gls{MI} (dashed black line) is our goal.}
	\label{fig:consistency}
\end{figure}
For the small room with Gaussian noise depicted in Fig.~\ref{fig:corr_4x4_gauss}, all \gls{MINE} models overestimate the \gls{MI} minimally, as previously seen in Fig.~\ref{fig:convergence_4x4_gauss}. The small models with the least capacity have slightly less correlation than the medium models, but come closest to the Monte Carlo \gls{MI} due to seemingly less overfitting. If the Gaussian noise is replaced with uniform noise in Fig.~\ref{fig:corr_4x4_uni}, small models show high variation, as indicated in Fig.~\ref{fig:convergence_4x4_uni}, while medium and large \gls{MINE} models give very consistent \gls{MI} approximations. For the large room with Gaussian noise in Fig.~\ref{fig:corr_10x10_gauss}, small overestimation but high correlation with the Monte Carlo \gls{MI} is achievable using medium and large models, while the small models underestimate the \gls{MI}, and have large sample spread. With uniform noise in Fig.~\ref{fig:corr_10x10_uni}, all models underestimate the \gls{MI}, and have larger spread. The large models come closest to approximating the Monte Carlo \gls{MI}, but they  have less correlation with it than the medium models. Some large models might not be fully converged yet after 700.000 epochs, as indicated in Fig.~\ref{fig:convergence_10x10_uni}. The \gls{MINE} models still give mostly monotonic \gls{MI} estimates (i.e. larger \gls{MI} estimates for larger ground truth \glspl{MI}), which is usually sufficient for optimization, as it allows to find better solutions.
\subsection{Pre-training \gls{MINE}}
\begin{figure}[h!]
	\centering
	\begin{subfigure}[b]{.24\textwidth}\
		\centering
		\includegraphics[width=\textwidth]
		{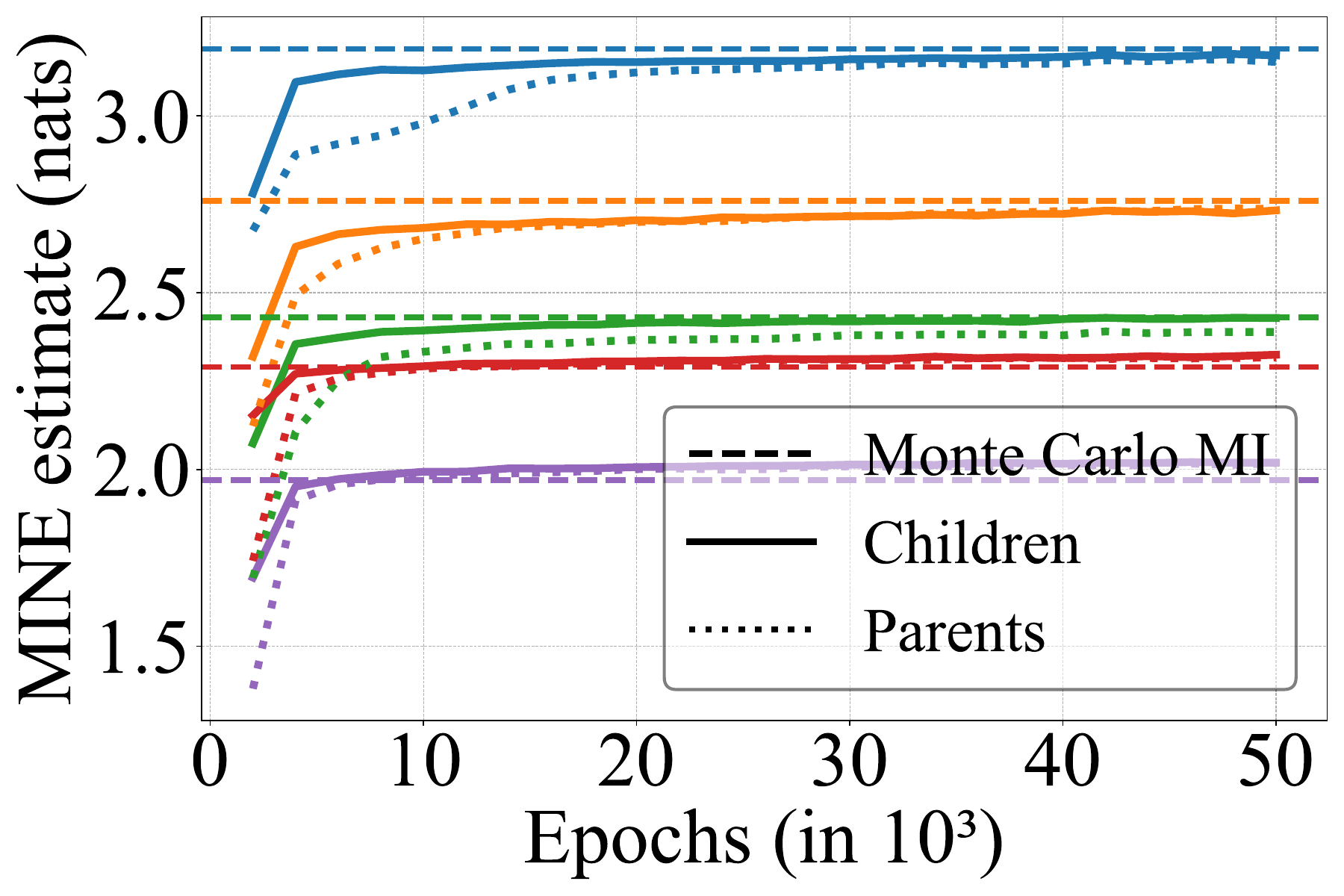}
		\caption{4x4 m room, Gaussian noise, small models.}
		\label{fig:pre_10x10_small}
	\end{subfigure}
	\hfill
	\begin{subfigure}[b]{.24\textwidth}
		\centering
		\includegraphics[width=\textwidth]{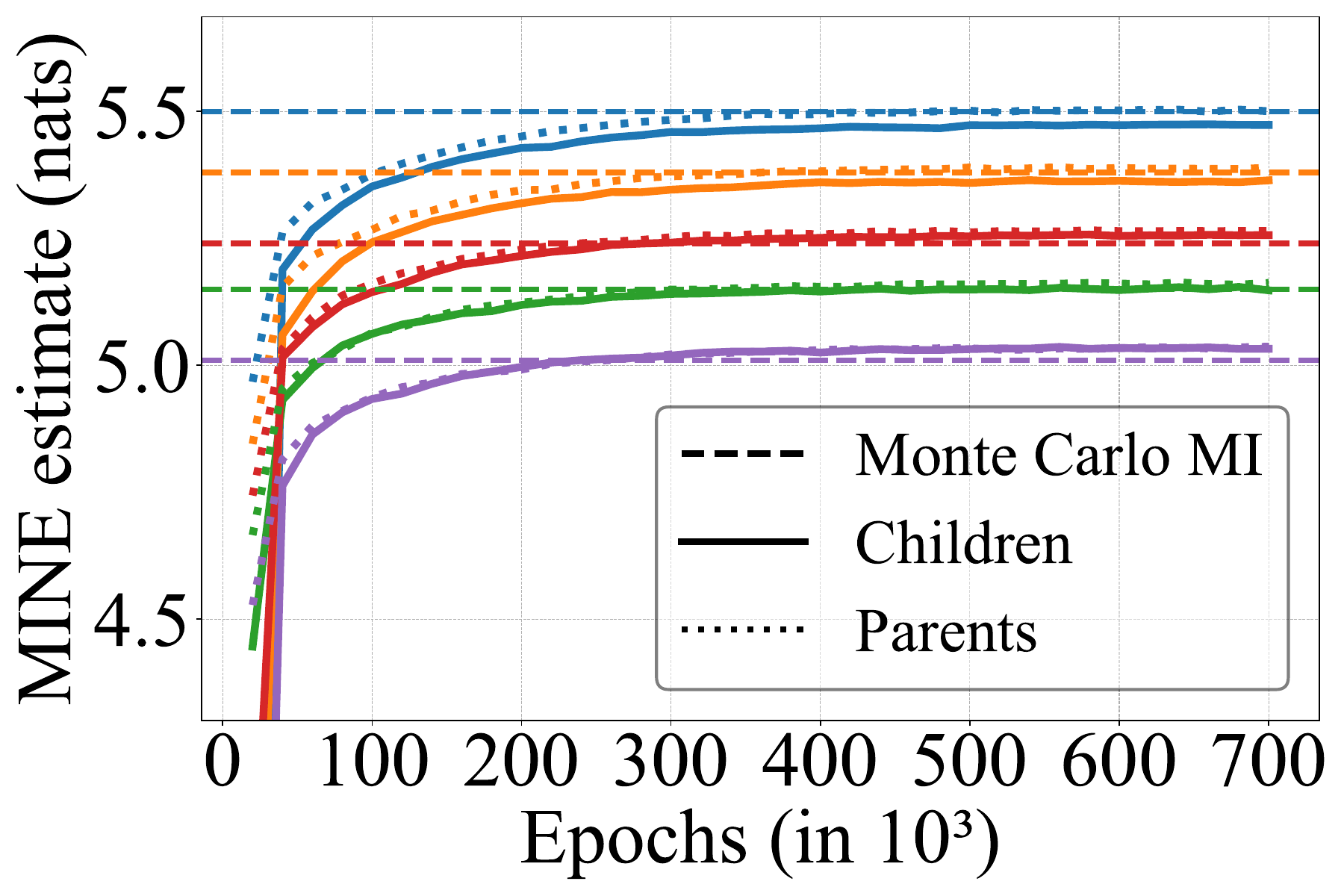}
		\caption{10x10 m room, Gaussian noise, medium models.}
		\label{fig:pre_10x10_medium}
	\end{subfigure}
	\caption{Learning curves of original parent \gls{MINE} models and of their children that are fine-tuned based on the pre-trained parents.}
	\label{fig:pretraining}
\end{figure}

\gls{MI} optimization requires fast \gls{MI} estimation, as this process has to be repeated for many reference placements. Therefore, we test if pre-trained \gls{MINE} models can be fine-tuned with data of similar reference placements instead of always training a new \gls{MINE} model from scratch. Five small and five medium \gls{MINE} models serve as pre-trained parents and five respective children \gls{MINE} models are fine-tuned based on those parents. Parents and children share the same reference placements. Each child selects its parent based on the smallest average Euclidean distance between its own position reference locations and the position reference locations of all parents, except the one with identical reference placement. To simplify this selection, we only compute the distances of references with the same identifiers. In \gls{MLAT} systems, reference identifiers can be flipped for equivalent placements. Integrating this permutation invariance of references for parent selection would require also finding optimal correspondences \cite{perm_ga}. For the medium models, we use the large room with Gaussian noise. For the small models, we choose the small room with Gaussian noise and average over only 2000 epochs for \gls{MI} estimation. The learning curves of the parents and children are given in Fig.~\ref{fig:pretraining}. The children use the same learning rate scheduler as the parents and start training from epoch zero.
The small children models in Fig.~\ref{fig:pre_10x10_small} show faster training and converge to the same \gls{MI} values as their parents. The medium parent and children models depicted in Fig.~\ref{fig:pre_10x10_medium} also converge to the same \gls{MI} values, however no gain in training times is observed. No deduction about the applicability of \gls{MINE} pre-training can be given for this task, as it appears to depend on the specific problem. As another technique to accelerate \gls{MINE} training, an early stopping mechanism, e.g. based on the standard deviation of the last $\hat{I}_{MINE}$ estimates, could be integrated.
\subsection{Mutual Information for \gls{MLAT}}
In this section, we investigate the usefulness of \gls{MI} for \gls{MLAT} systems. We simulate $D=1000$ noisy measurements per \gls{UE} location for Monte Carlo \gls{MI} estimation and \gls{RMSE} estimation with an \gls{MLAT} algorithm, using one set of position reference measurements per localization. Our \gls{MLAT} algorithm uses the $\mathcal{R}\subseteq \mathcal{L}$ detected references to estimate the 2D \gls{AMR} position $\underline{\hat{q}}=[\hat{x}, \hat{y}]^T\in\mathbb{R}^2$ with
\begin{align}
	\underline{\hat{q}} =
	\arg\min_{x,y} \sum_{j\in\mathcal{R}} \left( \left\| \underline{p}_j - [x,y,z_0]^T \right\|_2 - d_j \right)^2, \label{eq:mle}
\end{align}
where $d_j$ is the measured distance to reference $j$. Therefore, we first formulate the following linear system of equations  for the localization problem \cite{mlat} (here adapted for 2D localization) $\mathbf{A}\underline{\overline{x}} = \underline{b}$ with
\begin{equation}
\begin{aligned}
	\mathbf{A} &= 
	\begin{bmatrix}
		1 & -2x_1 & -2y_1 \\
		1 & -2x_2 & -2y_2 \\
		\vdots & \vdots & \vdots \\
		1 & -2x_{R} & -2y_{R}
	\end{bmatrix}\in\mathbb{R}^{R\times3},\quad\underline{\overline{x}}=
	\begin{bmatrix}
		d \\
		x \\
		y \\
	\end{bmatrix}\in\mathbb{R}^3,\\
	\underline{b}&=
	\begin{bmatrix}
		d_1^2 - x_1^2 - y_1^2 - z_1^2 -z_0^2 + 2z_0z_1\\
		d_2^2 - x_2^2 - y_2^2 - z_1^2 -z_0^2 + 2z_0z_1\\
		d_R^2 - x_R^2 - y_R^2 - z_1^2 -z_0^2 + 2z_0z_1\\
	\end{bmatrix}\in\mathbb{R}^R,
\end{aligned}
\end{equation}
and $d = x^2+y^2$.
We calculate an initial estimate  $\underline{\hat{x}} =  \mathbf{A}^+\underline{b}= [\hat{d}, \hat{x}, \hat{y}]^T$, where the Moore-Penrose pseudoinverse $\mathbf{A}^+\in\mathbb{R}^{3\times R}$ of $\mathbf{A}$ is computed by \gls{SVD} of $\mathbf{A}$. The initial estimates of $\hat{x}$ and $\hat{y}$ are then refined with nonlinear least squares using the Levenberg-Marquard algorithm.

\begin{figure}[h!]
	\centering
	\begin{subfigure}[b]{.24\textwidth}\
	\includegraphics[width=\textwidth]
	{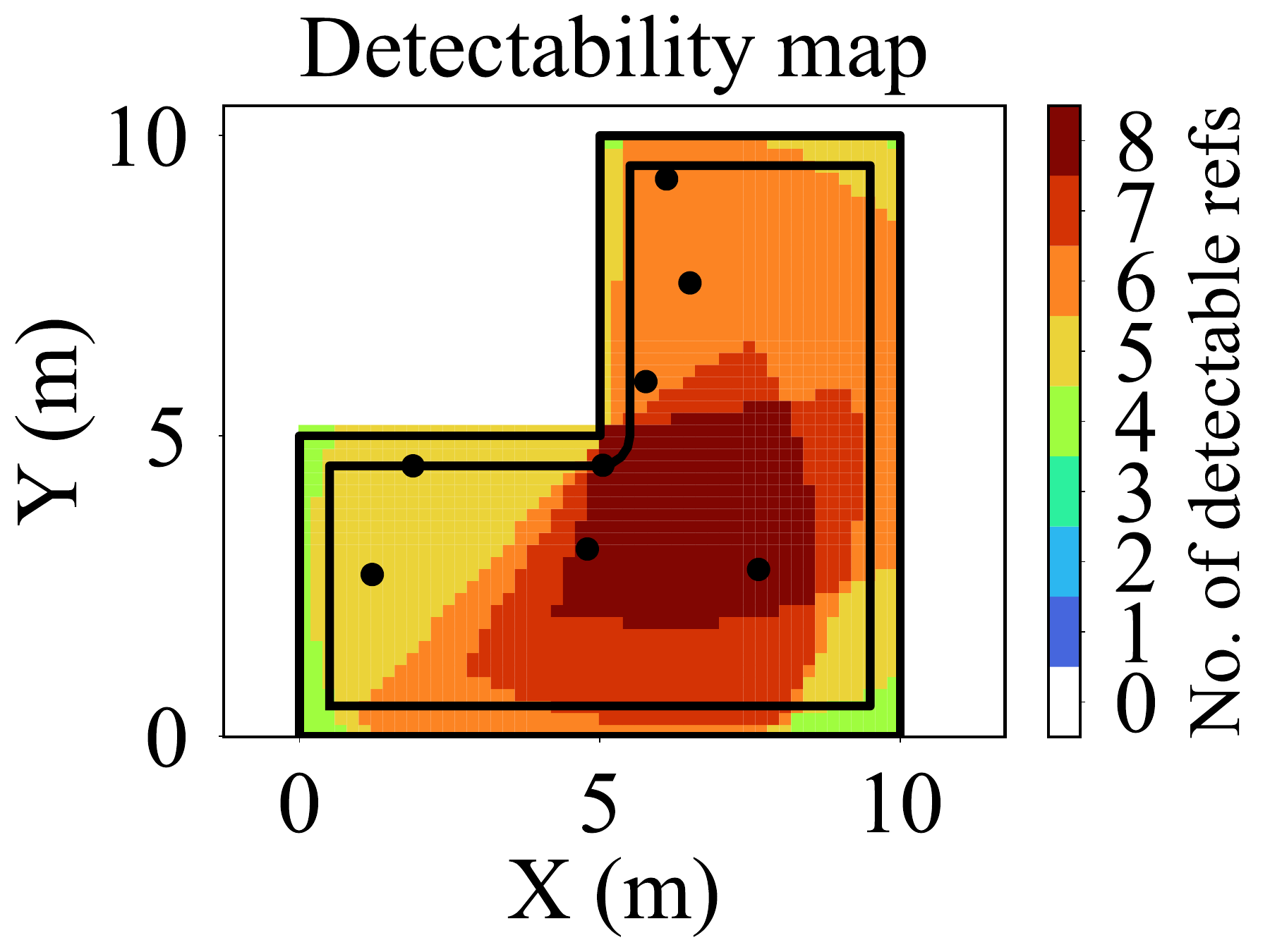}
	\caption{10x10 m room.}
	\label{fig:vis_rps}
	\end{subfigure}
	\hfill
	\begin{subfigure}[b]{.24\textwidth}\
		\includegraphics[width=\textwidth]
		{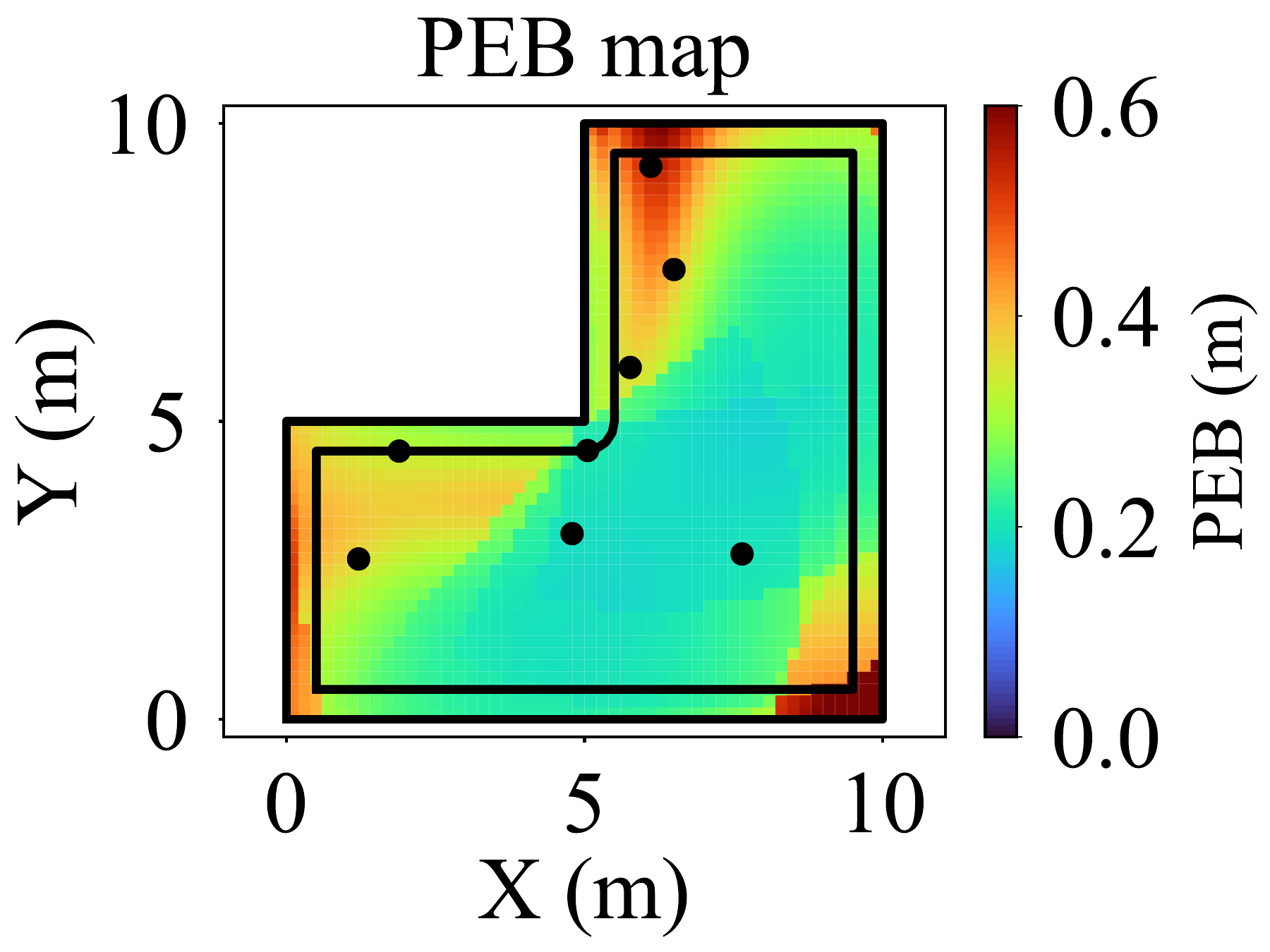}
		\caption{10x10 m room.}
		\label{fig:ideal_rmse}
	\end{subfigure}
	\vfill
	\begin{subfigure}[b]{.24\textwidth}
		\centering
		\includegraphics[width=\textwidth]{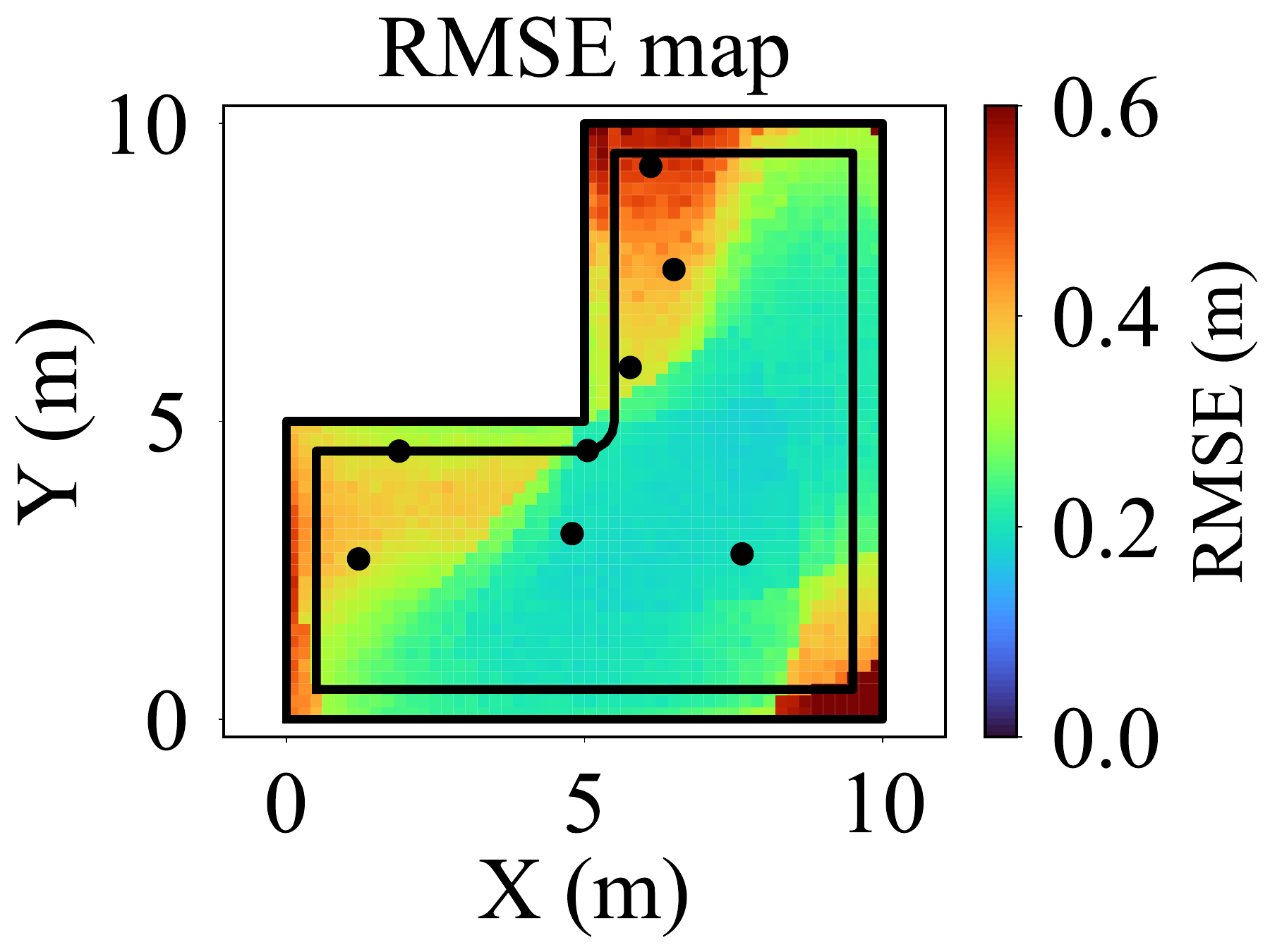}
		\caption{10x10 m room, Gaussian noise.}
		\label{fig:est_rmse_gauss}
	\end{subfigure}
	\hfill
	\begin{subfigure}[b]{.24\textwidth}\
		\centering
		\includegraphics[width=\textwidth]
		{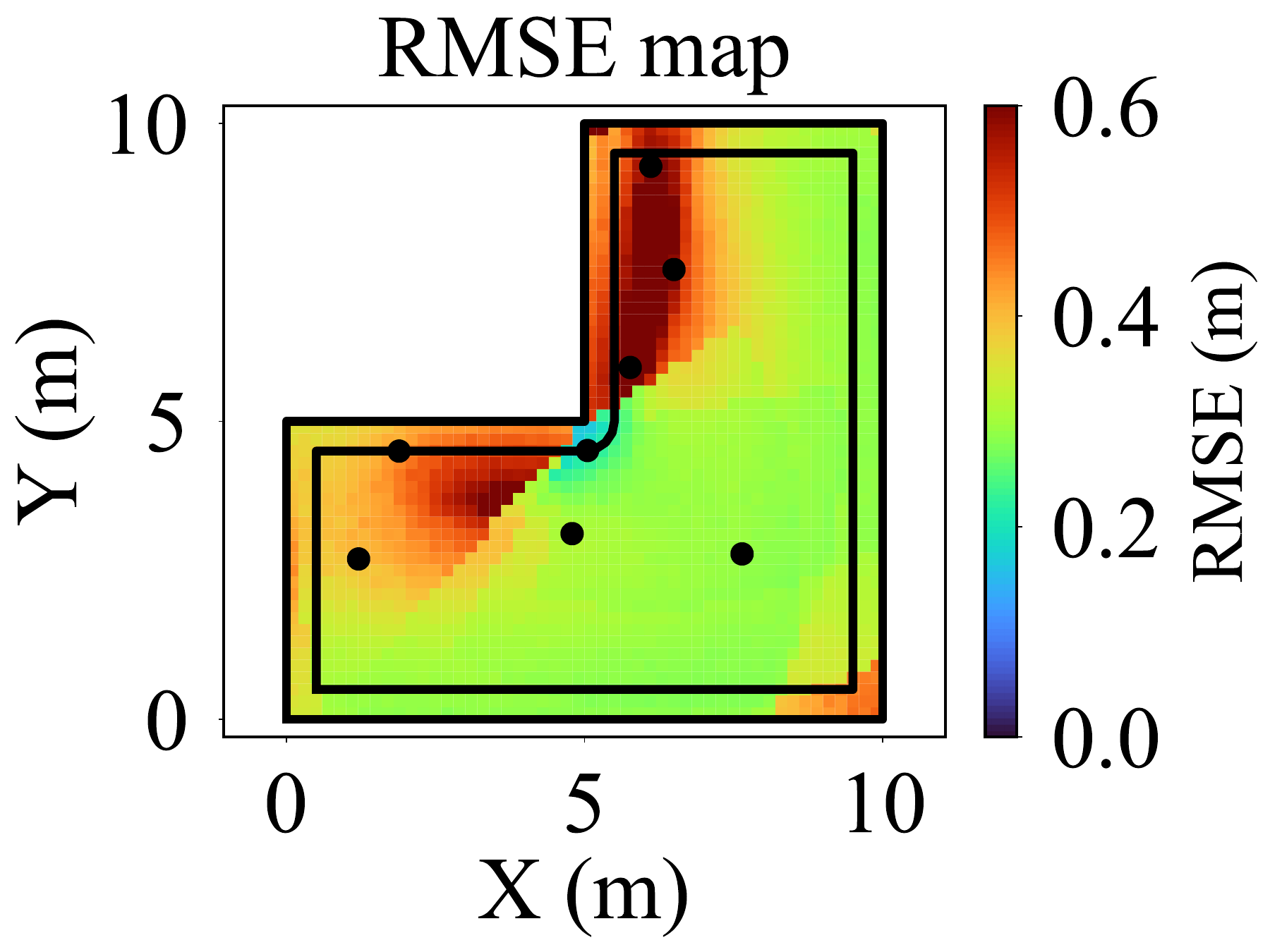}
		\caption{10x10 m room, uniform noise.}
		\label{fig:est_rmse_uni}
	\end{subfigure}
	\vfill
	\begin{subfigure}[b]{.24\textwidth}
		\centering
		\includegraphics[width=\textwidth]{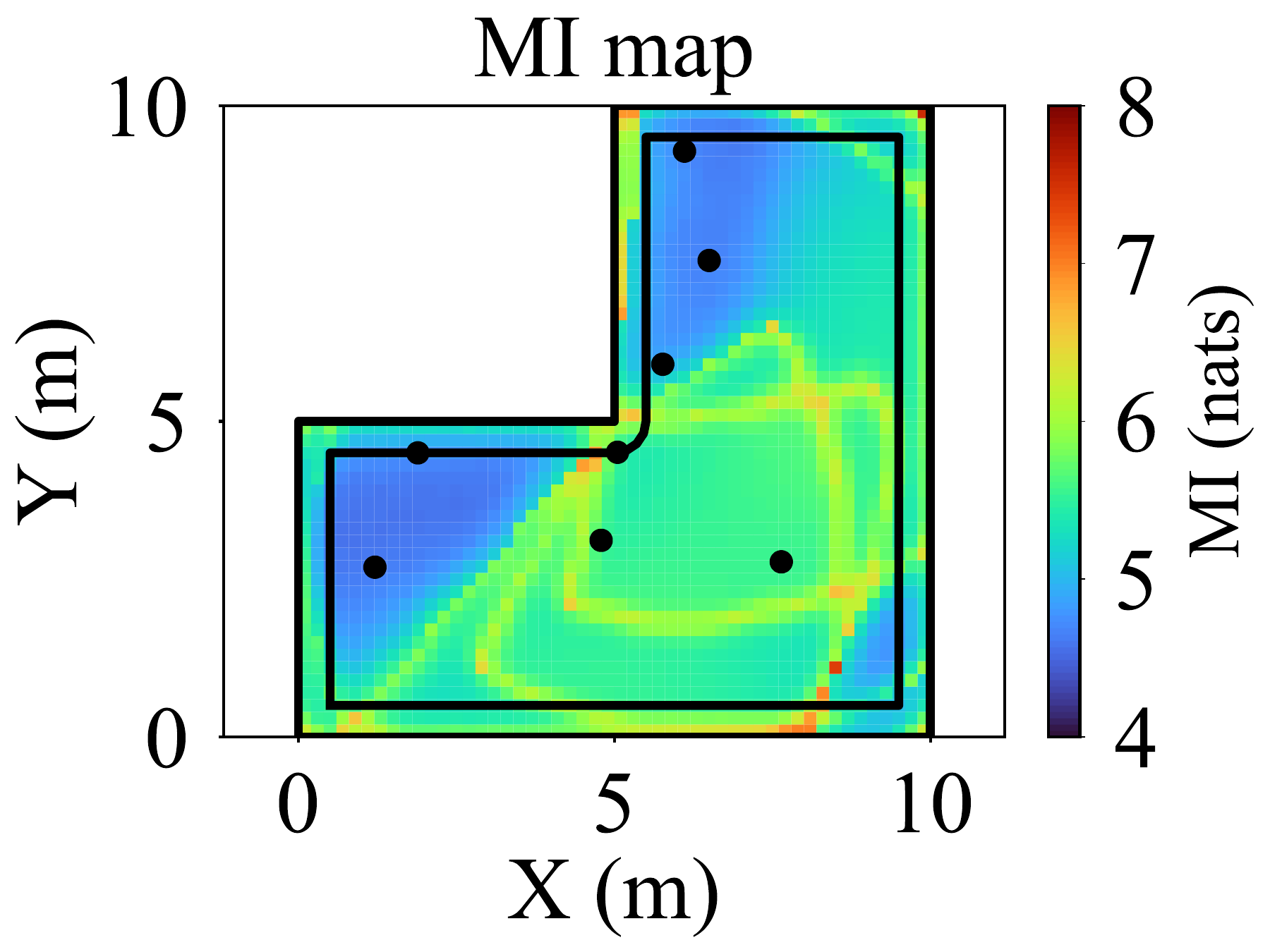}
		\caption{10x10 m room, Gaussian noise.}
		\label{fig:est_mi_gauss}
	\end{subfigure}
	\begin{subfigure}[b]{.24\textwidth}
	\centering
	\includegraphics[width=\textwidth]{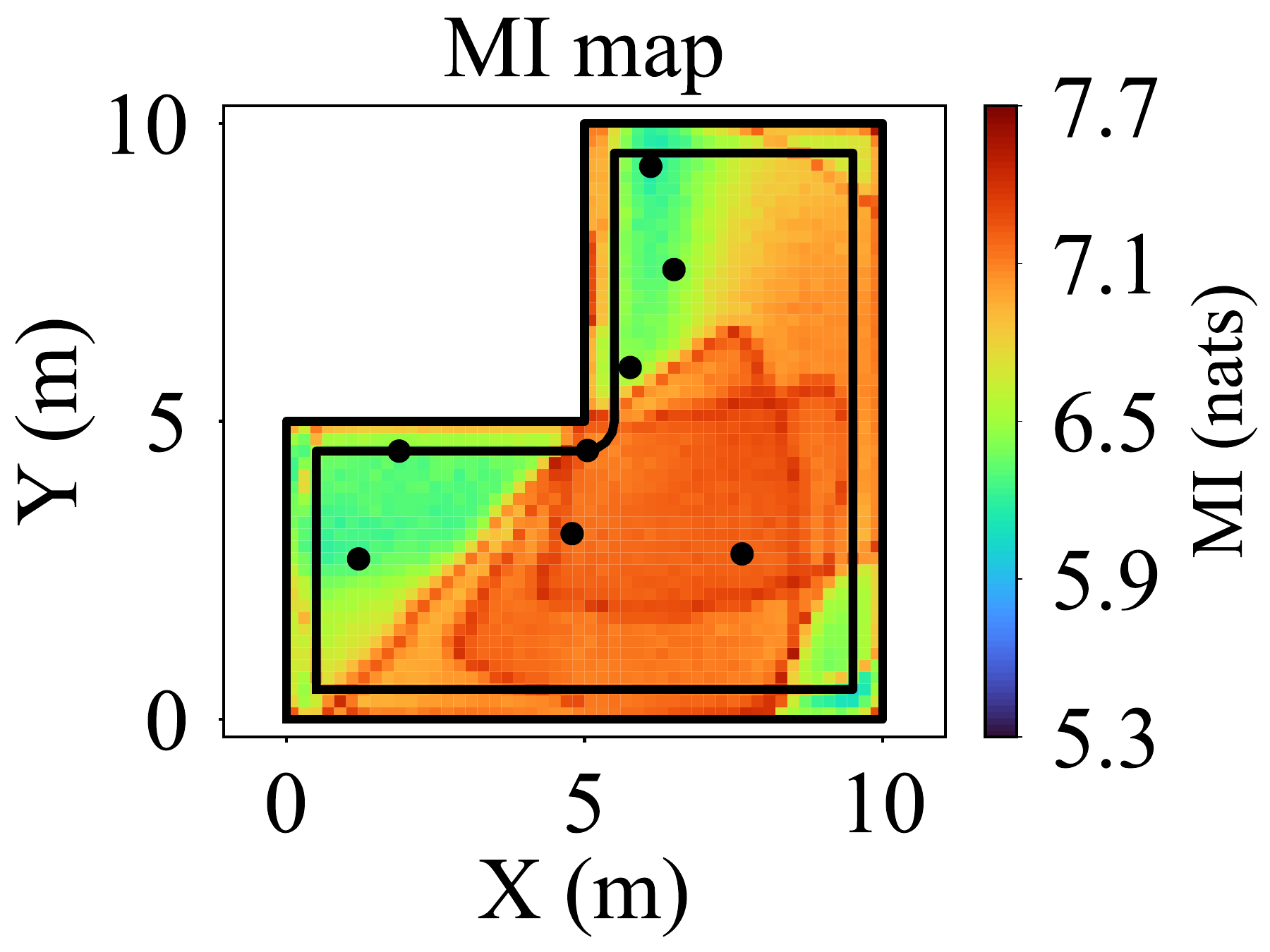}
	\caption{10x10 m room, uniform noise.}
	\label{fig:est_mi_uni}
	\end{subfigure}
	\caption{Comparison of the estimated \gls{RMSE} maps and the \gls{MI} maps from Monte Carlo simulation for both Gaussian and biased uniform noise. The black dots are the position references.}
	\label{fig:maps}
\end{figure}

This experiment is performed for a reference placement in the large room. The corresponding number of detectable references in Fig.~\ref{fig:vis_rps} and the \gls{PEB} from \eqref{eq:rmse_ideal} in Fig.~\ref{fig:ideal_rmse} serve as reference. For evaluation, we show the estimated \glspl{RMSE} from least squares \gls{MLAT} per grid element (averaged over all 1000 i.i.d. samples) for the Gaussian noise in Fig.~\ref{fig:est_rmse_gauss} and for the uniform noise in Fig.~\ref{fig:est_rmse_uni}. We use a biased uniform distribution $n_i^{j,o} \sim\mathcal{U}\left(0, 2\sigma_r\right)$ to evaluate larger deviations from the i.i.d., zero mean, Gaussian noise assumption in \eqref{eq:rmse_ideal}. Even though \gls{PEB} under given Gaussian measurement noise assumptions is a popular evaluation measure, the assumptions often don't perfectly hold in reality. Bias may e.g. come from erroneous calibration and uniform error distributions from quantization. For \gls{MI} estimation, we therefore change \eqref{eq:uni} to
\begin{align}
	p_{\text{Uni}} &=
	\begin{cases}
		\frac{1}{2 \sigma_r}, & \text{if } \left| m_{i'}^{j,o} - \sigma_r - r_i^j \right| \leq  \sigma_r \\
		0, & \text{otherwise}.
	\end{cases}
\end{align}

The \gls{MI} contributions of each \gls{UE} position estimated with Monte Carlo simulation (the MI contributions for each $x$ in \eqref{eq:MC_cond_entropy}) for the Gaussian noise are given in Fig.~\ref{fig:est_mi_gauss} and for the uniform noise in Fig.~\ref{fig:est_mi_uni}. 

As expected, the estimated \gls{RMSE} closely matches \gls{PEB} for identical Gaussian noise, as visible when comparing Fig.~\ref{fig:ideal_rmse} with Fig.~\ref{fig:est_rmse_gauss}. The largest deviations are found in the suboptimal region in the top of the room. For uniform noise in Fig.~\ref{fig:est_rmse_uni}, where the zero mean Gaussian noise assumption of \gls{PEB} is violated, the correlation decreases, but \gls{PEB} is still a good measure for \gls{MLAT} performance. \gls{MI} evaluated with Gaussian noise in Fig.~\ref{fig:est_mi_gauss} shows less, but still high negative correlation with the estimated \gls{RMSE}. However, large \gls{MI} values are visible where the number of detectable references changes, caused by missing \gls{LOS} (straight lines) or the limited sensing range (rings) of references. While \gls{PEB}/\gls{RMSE} values decrease with more available position references (for the \gls{FIM} based \gls{PEB}, having more available references is always better), the information on the change in detectable references increases \gls{MI} in those regions for grid elements with both more and fewer references. This should, however, be seen as an artifact of our simulator with deterministic detectability computation of references, which is usually stochastic by nature. In case of uniform noise in Fig.~\ref{fig:est_mi_uni}, \gls{MI} values are overall larger with smaller dynamic range, but correlation with the estimated \gls{RMSE} in Fig.~\ref{fig:est_rmse_uni} is high.

\begin{figure}[h!]
	\centering
	\begin{subfigure}[b]{.40\textwidth}\
		\centering
		\includegraphics[width=\textwidth]
		{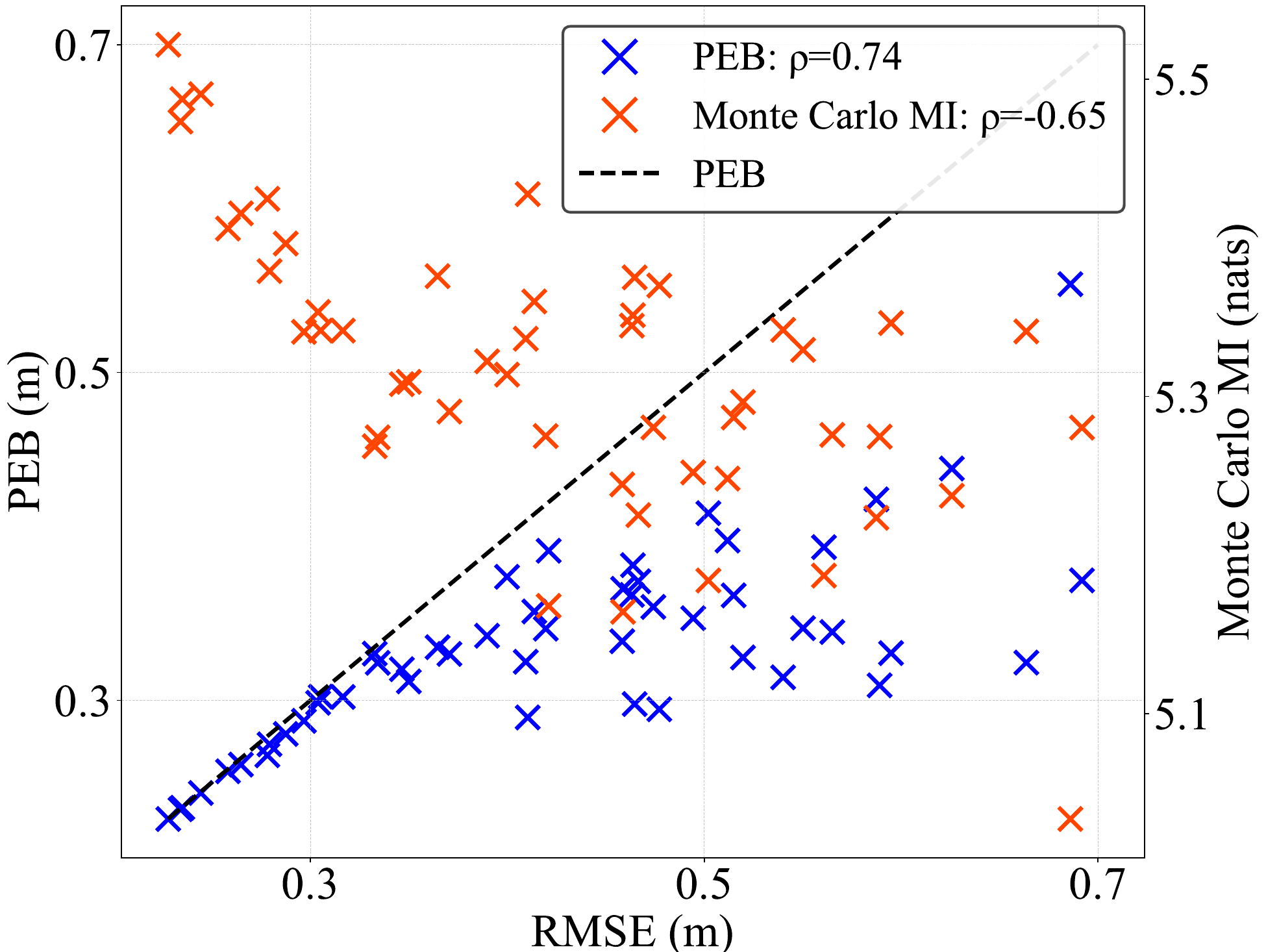}
		\caption{10x10 m room, Gaussian noise.}
		\label{fig:mlat_50_Gauss}
	\end{subfigure}
	\vfill
	\begin{subfigure}[b]{.40\textwidth}
		\centering
		\includegraphics[width=\textwidth]{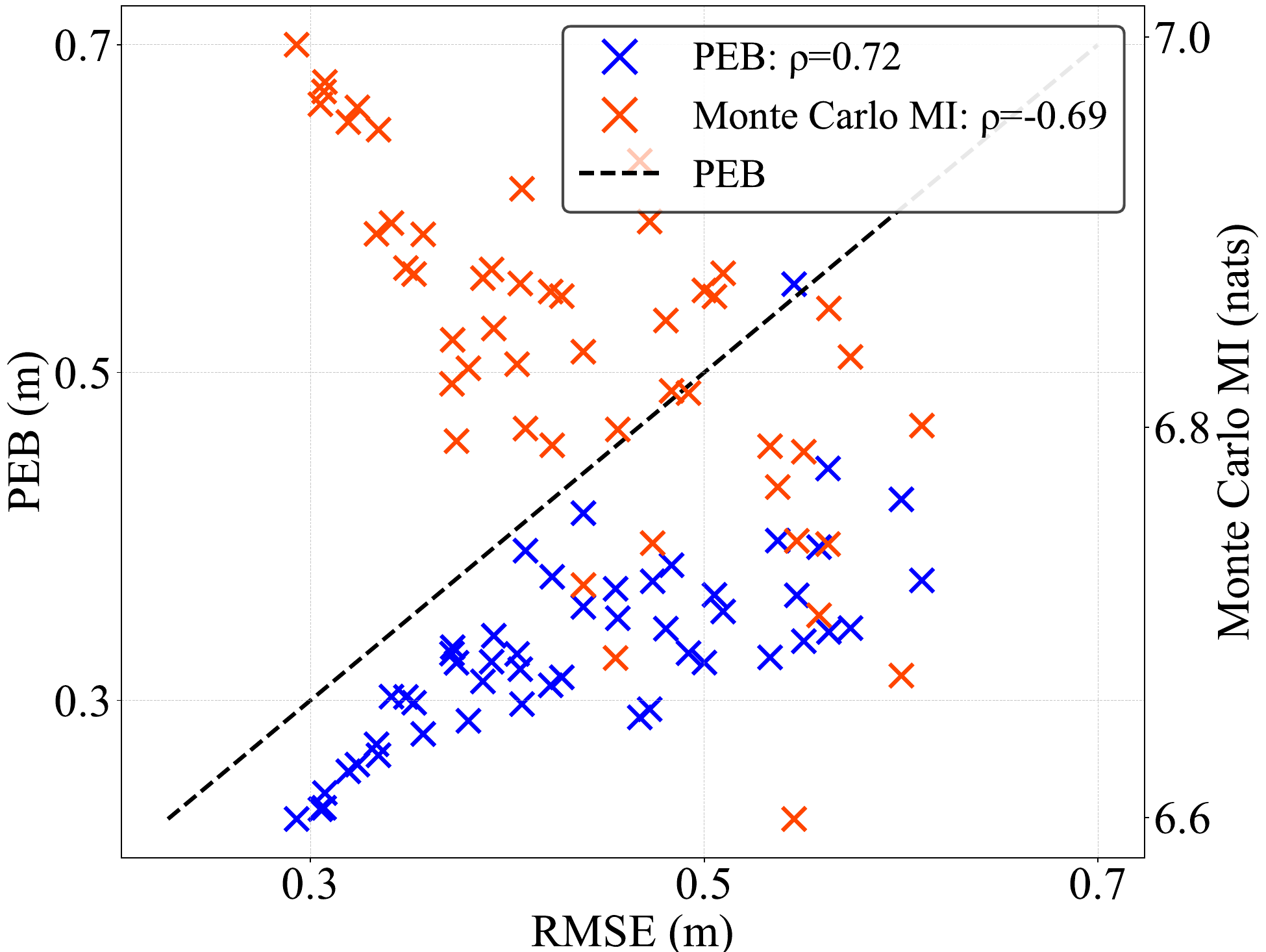}
		\caption{10x10 m room, uniform noise.}
		\label{fig:mlat_50_uniform}
	\end{subfigure}
	\caption{Comparison of downstream \gls{RMSE} performance with \gls{PEB} and estimated Monte Carlo \gls{MI}. The Pearson correlation coefficients $\rho$ between the estimated \gls{RMSE} through nonlinear least squares \gls{MLAT} and the respective measures \gls{PEB} and  Monte Carlo \gls{MI} are given.}
	\label{fig:mlat_50}
\end{figure}

To evaluate the relation of the estimated \gls{MLAT} \gls{RMSE} with the measures \gls{PEB} and Monte Carlo \gls{MI}, we reuse the 50 reference placements for the large room from Fig.~\ref{fig:corr_10x10_gauss} and Fig.~\ref{fig:corr_10x10_uni} and compute the averaged measures over all i.i.d. samples and grid elements per reference placement. The results for all 50 reference placements are shown in Fig.~\ref{fig:mlat_50}.

With Gaussian noise in Fig.~\ref{fig:mlat_50_Gauss}, the implemented least squares \gls{MLAT} algorithm approximates the \gls{PEB} for good position reference placements with low \gls{RMSE}. Therefore, \gls{PEB} correlates very well with \gls{MLAT} performance. Bad placements cause larger deviations. We believe this is amplified by using only one snapshot instead of multiple independent sets of measurements for localization. Our local Levenberg-Marquard solver requires good initial estimates and suffers from bad geometries, so this deviation from \gls{PEB} is expected and also visible in Fig.~\ref{fig:est_rmse_gauss}. With the uniform noise distribution in Fig.~\ref{fig:mlat_50_uniform}, \gls{PEB} is still a good proxy for downstream \gls{MLAT} performance. 

\gls{MI} shows high negative correlation with localization accuracy. While \gls{PEB} has less correlation with \gls{MLAT} performance when the underlying zero mean Gaussian assumption doesn't hold, as shown in Fig.~\ref{fig:mlat_50_uniform}, correlation with \gls{MI} interestingly increases. It is, however, still worse than the computationally much cheaper \gls{PEB}. The relation between \gls{RMSE} and \gls{MI} was also computed under the assumptions that we have perfect knowledge about the measurement error distribution for \gls{MI} estimation and that this distribution doesn't change between \gls{MI} estimation and inference, which is quite unrealistic.
We deduct that while \gls{MI} shows high correlation with downstream localization performance and it might be a useful measure for more realistic and complex localization systems, or when different noise distributions or distortions are involved, it appears to not be an ideal choice for evaluating and optimizing the simple, simulated \gls{MLAT} system in this work.

\section{Conclusion}
We evaluated a neural \gls{MI} estimator called \gls{MINE} for the assessment of position reference placements of localization systems. \gls{MINE} shows better \gls{MI} approximations than previous non-neural \gls{MI} estimators and, as a data-driven method, it's generally applicable, requiring only sets of \gls{UE} positions and corresponding measurements for \gls{MI} estimation.  

Under Gaussian measurement noise assumptions, \gls{MI} approximation with \gls{MINE} was mostly successful, with slight \gls{MI} overestimations. With uniform noise which is not smooth, unlike the Gaussian, \gls{MI} estimation has proven to be difficult and consistent \gls{MI} approximations were only possible in the smaller evaluated room. In the more difficult, larger room, large statistics networks underestimated \gls{MI} the least. Computational times with the \gls{MINE} estimator are still long, which makes \gls{MI} approximation and, even more so, \gls{MI} optimization with \gls{MINE} costly. For our simulated \gls{MLAT} system, the much cheaper \gls{PEB} has shown better correlation with the accuracy of a \gls{MLAT} localization algorithm, even when the underlying Gaussian noise assumptions for \gls{PEB} are violated.

Benchmarking different \gls{MI} estimators, investigating other information measures, and applying more complex localization approaches pose interesting research directions for future work. Further, real-world datasets would give more insight into the practical use of \gls{MI} and \gls{MI} estimators for localization.
\section{Acknowledgements}
We want to thank our colleagues Louisa Fay for her previous MINE work at our institute, thus giving us the idea of applying MINE for localization, and Nico Reick and Ren\'{e} Tröger for consistently updating and maintaining our server, thereby enabling this work.
\bibliographystyle{ieeetr}
\bibliography{ref}
\end{document}